\documentclass[USenglish,onecolumn]{article}

\usepackage[utf8]{inputenc}
\usepackage[big]{dgruyter}
\baretabulars
\usepackage{enumerate}

\usepackage{natbib}
\usepackage{booktabs}
\usepackage{graphicx}

\usepackage{amsxtra}     
\usepackage{amsthm}
\usepackage{amssymb}
\usepackage{amsfonts}
\usepackage{graphicx}    
\usepackage{rotating}
\usepackage{color}
\usepackage{epsfig}
\usepackage{subfigure}   
\usepackage{verbatim}
\usepackage{textcomp}
\usepackage{multirow}
\usepackage{booktabs}
\usepackage{tabularx}
\usepackage{lscape}
\usepackage[font=small,labelfont=bf]{caption}
\usepackage[shortlabels]{enumitem}
\usepackage[table,xcdraw]{xcolor}
\usepackage{array,booktabs,mathtools}
\usepackage[section]{placeins}
\usepackage{comment}

\newcommand{\beginsupplement}{%
        \setcounter{table}{0}
        \renewcommand{\thetable}{A\arabic{table}}%
        \setcounter{figure}{0}
        \renewcommand{\thefigure}{B\arabic{figure}}%
     }
     
\newcommand{\condind}{\mathrel{\text{\scalebox{1.07}{$\perp\mkern-10mu\perp$}}}}

\DeclareMathOperator*{\argmin}{arg\,min}
  
\begin{document}

  \articletype{Research Article{\hfill}Open Access}

  \author[1]{Youfei Yu}
  
  \author[2]{Jiacong Du}

  \author*[3]{Min Zhang}
  
  \author[4]{Zhenke Wu}
  
  \author[5]{Andrew M. Ryan}
  
  \author[6]{Bhramar Mukherjee}

  \affil[1]{Department of Biostatistics, University of Michigan; E-mail: youfeiyu@umich.edu}
  
  \affil[2]{Department of Biostatistics, University of Michigan; E-mail: jiacong@umich.edu}

  \affil[3]{Department of Biostatistics, University of Michigan; E-mail: mzhangst@umich.edu}
  
  \affil[4]{Department of Biostatistics, University of Michigan; E-mail: zhenkewu@umich.edu}
  
  \affil[5]{Department of Health Management and Policy, University of Michigan; E-mail: amryan@umich.edu}

  \affil[6]{Department of Biostatistics, University of Michigan; E-mail: bhramar@umich.edu}
  
  \title{\huge Outcome Adaptive Propensity Score Methods for Handling Censoring and High-Dimensionality: Application to Insurance Claims}

  \runningtitle{Outcome-adaptive propensity score methods}


  \begin{abstract}
{Propensity scores are commonly used to reduce the confounding bias in non-randomized observational studies for estimating the average treatment effect. An important assumption underlying this approach is that all confounders that are associated with both the treatment and the outcome of interest are measured and included in the propensity score model. In the absence of strong prior knowledge about potential confounders, researchers may agnostically want to adjust for a high-dimensional set of pre-treatment variables. As such, variable selection procedure is needed for propensity score estimation. In addition, recent studies show that including variables related to treatment only in the propensity score model may inflate the variance of the treatment effect estimates, while including variables that are predictive of only the outcome can improve efficiency. 
In this paper, we propose a flexible approach to incorporating outcome-covariate relationship in the propensity score model by including the predicted binary outcome probability (OP) as a covariate. Our approach can be easily adapted to an ensemble of variable selection methods, including regularization methods and modern machine learning tools based on classification and regression trees. 
We evaluate our method to estimate the treatment effects on a binary outcome, which is possibly censored, among multiple treatment groups.
Simulation studies indicate that incorporating OP for estimating the propensity scores can improve statistical efficiency and protect against model misspecification. The proposed methods are applied to a cohort of advanced stage prostate cancer patients identified from a private insurance claims database for comparing the adverse effects of four commonly used drugs for treating castration-resistant prostate cancer.}
\end{abstract}

  \journalname{Journal of Causal Inference}

  \journalyear{2019}
  \journalvolume{1}

\maketitle
\section{Introduction}

To obtain an unbiased estimate for the causal average treatment effect (ATE) using data from observational studies, it is important to control for confounding by adjusting for the differences in pre-treatment baseline covariates between treatment groups. A common tool for confounder adjustment is the propensity score, defined as the conditional probability of receiving a specific treatment given a set of covariates \citep{rosenbaum1983}, which is usually unknown in observational studies and needs to be estimated from the data. The estimated propensity scores can be used in matching \citep{stuart2010, yang2016}, weighting \citep{lunceford2004}, regression adjustment \citep{zhou2019}, among other methods, for estimating the causal treatment effect. Propensity score-based causal estimators rely on robust and accurate estimation for the propensity scores.

There has been a large body of work focusing on the estimation and use of propensity scores in a low-dimensional setting, where the sample size is far greater than the number of candidate covariates. In this setting, empirical researchers often estimate the propensity scores by fitting a logistic regression model for binary treatment, or a multinomial logistic regression model for more than two treatment groups. In the era of `big data', large healthcare databases are increasingly being used to conduct comparative effectiveness research. For example, insurance claims data were used to compare the safety of four drugs on the market prescribed for patients with metastatic castration-resistant prostate cancer \citep{yu2021}. A massive collection of candidate covariates, such as demographics, socioeconomic status, clinical measurements, and diagnosis codes, can be ascertained from patients' health care or insurance claims data, and the number of covariates may exceed the sample size in each treatment group. Standard parametric regression models may become problematic when the dimensionality increases, and a key challenge is to identify variables to be included in the propensity score model from a high-dimensional set of measured covariates. There has been considerable interest in developing methods that perform variable selection for estimating propensity/balancing scores in the high-dimensional setting. For instance, \citet{schneeweiss2009} proposed the high-dimensional propensity score (hd-PS) algorithm that selects covariates to facilitate high-dimensional propensity score adjustment using health care claims data. Specifically, they rank each covariate based on its potential for controlling confounding by assessing the covariate's mean and univariate association with the treatment and outcome, and then pick the top $k$ covariates for inclusion in the propensity score modeling. 

Parametric models require variable selection and specification of functional forms. A practical difficulty is that misspecification of the model can result in substantial bias of the estimated treatment effect \citep{kang2007}. Machine learning methods are well-known for their powerful predictive performance and ability to handle complex and nonlinear relationship between outcomes and covariates. There has been a growing interest in using machine learning techniques for propensity score modeling \citep{setoguchi2008, lee2010, ju2019a, tu2019a}. \citet{setoguchi2008} compared several data mining techniques that optimize the prediction of treatment status, including classification and regression trees (CART), pruned CART, and neural networks, in the context of propensity score matching with a continuous outcome. \citet{lee2010} extended the work of \citet{setoguchi2008} to the setting with a binary outcome, and evaluated the performance of CART, pruned CART, bootstrap aggregated (bagged) CART, random forests, and boosted CART with regard to propensity score weighting. Both works focused on the low-dimensional setting, and demonstrated that machine learning methods, such as CART and neural networks, are promising alternatives to parametric modeling for the estimation of propensity scores in the presence of nonadditivity and/or nonlinearity in the true treatment assignment mechanism. However, numerical evaluations of these machine learning techniques for high-dimensional covariates and multiple treatment groups with a binary endpoint remain limited.

These aforementioned approaches only consider the treatment-covariate relationship when modeling the propensity/balancing scores, and fail to incorporate the outcome models into the treatment modeling process. Studies illustrate that using covariates that are associated with the treatment but not the outcome will inflate the variance of the estimators of ATE without reducing the bias \citep{deluna2011,shortreed2017}. On the other hand, adding covariates explaining the outcome but not the treatment to the propensity score model can improve the precision of the treatment effect estimates \citep{brookhart2006,shortreed2017}. These findings suggest that a highly predictive model for treatment assignment will not necessarily lead to the most efficient estimators of treatment effects. Therefore, standard variable selection methods designed for prediction, which rely only on the relationship between treatment and covariates, may yield sub-optimal results in the context of causal inference. There is an expanding literature on variable selection methods for causal inference that account for the information about the outcome-covariate relationship. \citet{shortreed2017} proposed the Outcome-Adaptive LASSO (OAL) method, which adopts the adaptive LASSO framework \citep{zou2006} and places smaller adaptive weights on the outcome predictors in the propensity score model to favor the inclusion of outcome-predictive covariates. As a result, heavier penalties are imposed on variables relevant to treatment only than variables predictive of outcome only. Other works that allow the outcome information to contribute to variable selection for propensity score modeling include Outcome Highly Adaptive LASSO proposed by \citet{ju2020} and Bayesian Adjustment for Confounding proposed by \citet{zigler2014}. However, how the information about the outcome-covariate relationship can be incorporated into tree-based machine learning methods, such as CART, and to what extent can outcome models contribute to the efficiency gain in high-dimensional data settings with multiple treatment groups have been less studied.

In this paper, we propose a flexible approach to incorporating outcome-covariate relationship in the propensity score model by including the predicted outcome probability (OP) as a covariate. Our approach can be easily adapted to an ensemble of variable selection methods, including regularization methods and tree-based machine learning algorithms. We evaluate our approach for estimating the causal average treatment effect with multiple treatment groups and high-dimensional covariates. We use the inverse probability weighting (IPW) estimator, although the estimated propensity scores considered in this study are applicable to any propensity score-based methodology, such as propensity score matching. We also aim to account for censoring at the same time, where the bias due to censoring is controlled for by applying the inverse probability of remaining uncensored as weights to the outcome. We focus on estimating the treatment effects on a binary outcome (that is possibly censored) among multiple treatment groups. To the best of our knowledge, the literature lying within the intersection of high-dimensionality, complexity of the associations between treatment and covariates, multiple treatments, censored outcomes, and outcome-adaptive propensity score modeling is largely absent. 

In Section~\ref{s:Chap3_definition}, we introduce the notation and basic setup of the problem. Section~\ref{s:Chap3_selection} describes the algorithm for selecting variables to be included in the treatment model. Section~\ref{s:Chap3_choices} outlines the methods considered for the final treatment model that is used to estimates the propensity scores. Section~\ref{s:Chap3_simulation} presents simulation studies that compare the methods in settings with high-dimensional covariates and potentially complex underlying treatment model (i.e., model with nonlinearity and/or nonadditivity). We also consider a setting where censoring exists. The simulation results illustrate the efficiency gain provided by leveraging the information about the outcome-covariate relationship when estimating the propensity scores. Section~\ref{s:Chap3_data} presents an example that compares the risks of possibly right-censored hospitalization and emergency room (ER) visits of four prostate cancer treatments using data from an insurance claims database. The binary outcomes (i.e., whether the patient was hospitalized/admitted to ER within 180 or 360 days of treatment initiation) in this data example were subject to censoring, as the follow-up period tended to terminate early due to drop-out or treatment switch. We close with a brief discussion.

\section{Definition of the Problem and Notation}
\label{s:Chap3_definition}
\subsection{Notation and Assumptions}
We consider $n$ independent individuals, indexed by $i$, with $\boldsymbol{X}_i$ being a $p$-dimensional vector of covariates measured prior to receiving the treatment $Z_i$, where $Z_i=j\in\{1,\cdots,J\}$. We let $\mathcal{C}$, $\mathcal{Z}$, $\mathcal{Y}$, and $\mathcal{S}$ be the indices of confounders (i.e., covariates associated with both treatment and outcome), covariates predictive of treatment only, covariates predictive of outcome only, and covariates unrelated to both treatment and outcome (i.e., spurious covariates), respectively. Suppressing the index by $i$, $\boldsymbol{X}_\mathcal{C}$, $\boldsymbol{X}_\mathcal{Z}$, $\boldsymbol{X}_\mathcal{Y}$, and $\boldsymbol{X}_\mathcal{S}$ are mutually exclusive and $\boldsymbol{X}=\boldsymbol{X}_\mathcal{C}\cup\boldsymbol{X}_\mathcal{Z}\cup\boldsymbol{X}_\mathcal{Y}\cup\boldsymbol{X}_\mathcal{S}$. We further let $|\mathcal{C}|$, $|\mathcal{Z}|$, $|\mathcal{Y}|$, and $|\mathcal{S}|$ denote the cardinality of the corresponding set. Re-introducing subject index $i$, we let $T_i$ be the underlying time to the first event of interest for each individual, and $C_i$ be the censoring time. The outcome of interest is whether the event of interest occurs before a prespecified time point $d$, defined by $Y_i=I(T_i<d)$, which results in a possibly censored binary outcome. The information on $Y_i$ may not be completely available due to dropout, study termination, or treatment switch. In the absence of censoring, $T_i$ (and therefore $Y_i$) would be observed for all individuals, and the set of complete data is then $(\boldsymbol{X}_i,Z_i,Y_i)$. When the outcome variable is subject to right-censoring, $Y_i$ is observed only if the individual has not been censored before $d$, and we let $R_i=I\{C_i\ge\min(T_i,d)\}$ be the indicator of observing $Y_i$. 

Under the potential outcome framework \citep{rubin1974}, each individual is associated with a set of potential outcomes $\{Y^{(1)},\cdots,Y^{(J)}\}$, where $Y^{(j)}$ denotes the potential outcome had the individual received treatment $j$. The causal parameter of interest is the marginal ATE on the outcome between $j$ and $j'$, denoted by $\tau(j,j')=E\{Y^{(j')} \}-E\{Y^{(j)}\}$. The following assumptions are required for statistical identification of causal effects using the observable data \citep{yu2022inverse}:

\begin{enumerate}[I.]
    \item (\textit{Random sampling}) The individuals in the study are randomly sampled from the population;
    \item (\textit{Stable Unit Treatment Value Assumption, or SUTVA}) For any individual $i$, $i=1,\cdots,n$, if $Z_i=j$, then $Y_i=Y_i^{(j)}$, for all $j=1,\cdots,J$;
    \item (\textit{Unconfoundedness}) $\{Y_i^{(1)},\cdots, Y_i^{(J)}\}\condind Z_i|\boldsymbol{X}_i$; \label{assump:unconfound}
    \item (\textit{Overlap}) For all values of $j$ and $\boldsymbol{x}$, $0<\pi_j(\boldsymbol{x})<1$, where $\pi_j(\boldsymbol{x})=pr(Z_i=j|\boldsymbol{x})$;
    \item (\textit{Censoring at random}) $ C_i\condind\{T_i^{(1)},\cdots,T_i^{(J)}\}\Big|(Z_i, \boldsymbol{X}_i)$. \label{assump:censor}
\end{enumerate}

\subsection{Underlying Models for Outcome, Treatment, and Censoring, and Estimators for Average Treatment Effects}
We assume that the true outcome model for $Z_i=j$ is a logistic regression model,
\begin{align*}
    \text{logit}\,P(Y_i=1|\boldsymbol{W}_i, Z_i=j)=\boldsymbol{W}_i^T\boldsymbol{\beta}_j, 
\end{align*}
where $\boldsymbol{W}_i$ is a $p_w$-dimensional function of $(\boldsymbol{X}_\mathcal{C}^T, \boldsymbol{X}_\mathcal{Y}^T)^T$. The treatment assignment mechanism is governed by a multinomial logistic regression
\begin{align*}
    \log\frac{P(Z_i=j|\boldsymbol{V}_i)}{P(Z_i=J|\boldsymbol{V}_i)}=\boldsymbol{V}_i^T\boldsymbol{\alpha}_j,\quad j=1,\cdots,J,
\end{align*}
where $J$ is the reference level and $\boldsymbol{V}_i$ is a $p_v$-dimensional function of $(\boldsymbol{X}_\mathcal{C}^T, \boldsymbol{X}_\mathcal{Z}^T)^T$. Both $\boldsymbol{W}_i$ and $\boldsymbol{V}_i$ may contain nonlinear terms and interactions. We assume that $|\mathcal{C}|+|\mathcal{Y}|\ll p$ and $|\mathcal{C}|+|\mathcal{Z}|\ll p$, where $|\mathcal{C}|$, $|\mathcal{Y}|$, and $|\mathcal{Z}|$ are the numbers of variables in $\boldsymbol{X}_\mathcal{C}$, $\boldsymbol{X}_\mathcal{Z}$, and $\boldsymbol{X}_\mathcal{Y}$, respectively. Note that in practice, $\boldsymbol{X}_\mathcal{C}$, $\boldsymbol{X}_\mathcal{Y}$, and $\boldsymbol{X}_\mathcal{Z}$ are generally unknown and a common practice is to include all candidate covariates in the model. In the case where the ratio of $p$ to $n$ is relatively large, traditional regression models based on maximum likelihood estimation may not converge, and some variable selection procedure is required for model fitting. 


With respect to censoring, we assume a proportional hazard model for treatment $j=1,\cdots,J$,
\begin{align*}
    \lambda_j(t|\boldsymbol{U}_i)=\lambda_{0j}(t)\exp(\boldsymbol{U}_i^T\boldsymbol{\gamma}_j) , 
\end{align*}
where $\lambda_{0j}(t)$ is the treatment-specific baseline hazard function and $\boldsymbol{U}_i$ is a $p_u$-dimensional function of $\boldsymbol{X}_i$. The censoring model is assumed to have a moderate number of predictors ($p_u\ll n$) that are known \textit{a priori}. 

We estimate the ATE using the IPW estimator, 
\begin{align*}
    \hat{\tau}(j,j')=\frac{\sum_{i=1}^n\hat{w}_{i}I(Z_i=j')Y_i}{\sum_{i=1}^n\hat{w}_{i}I(Z_i=j')}\,-\, \frac{\sum_{i=1}^n\hat{w}_{i}I(Z_i=j)Y_i}{\sum_{i=1}^n\hat{w}_{i}I(Z_i=j)} ,
\end{align*}
where $\hat{w}_i=\sum_{j=1}^J 1/\hat{\pi}_j(\boldsymbol{X}_i)$ and the propensity scores $\hat{\pi}_j(\boldsymbol{X}_i)$, $j=1\cdots,J$, are required to be estimated from the data in an observational study. In the presence of censoring, the outcome is further weighted by the inverse probability of remaining uncensored at $d$,
\begin{align*}
    \hat{\tau}(j,j')=\frac{\sum_{i=1}^n\hat{w}_i^{\ast}R_iI(Z_i=j')Y_i}{\sum_{i=1}^n\hat{w}_i^{\ast}R_iI(Z_i=j')}\,-\, \frac{\sum_{i=1}^n\hat{w}_i^{\ast}R_iI(Z_i=j)Y_i}{\sum_{i=1}^n\hat{w}_i^{\ast}R_iI(Z_i=j)}.
\end{align*}
The weights $\hat{w}_i^{\ast}=\sum_{j=1}^J(\hat{\pi}_j(\boldsymbol{X}_i)\exp{\{\Lambda_{ij}(\min(T_i,C_i,d))\}})^{-1}$ \citep{yu2022inverse}, where $\Lambda_{ij}(t)$ is the cumulative hazard function of $C_i$ at $t$ for treatment $j$.

\section{Variable Selection for Dimensionality Reduction for the Propensity Score Model}
\label{s:Chap3_selection}

\begin{figure}[h]
    \centering
    \includegraphics[width=\textwidth]{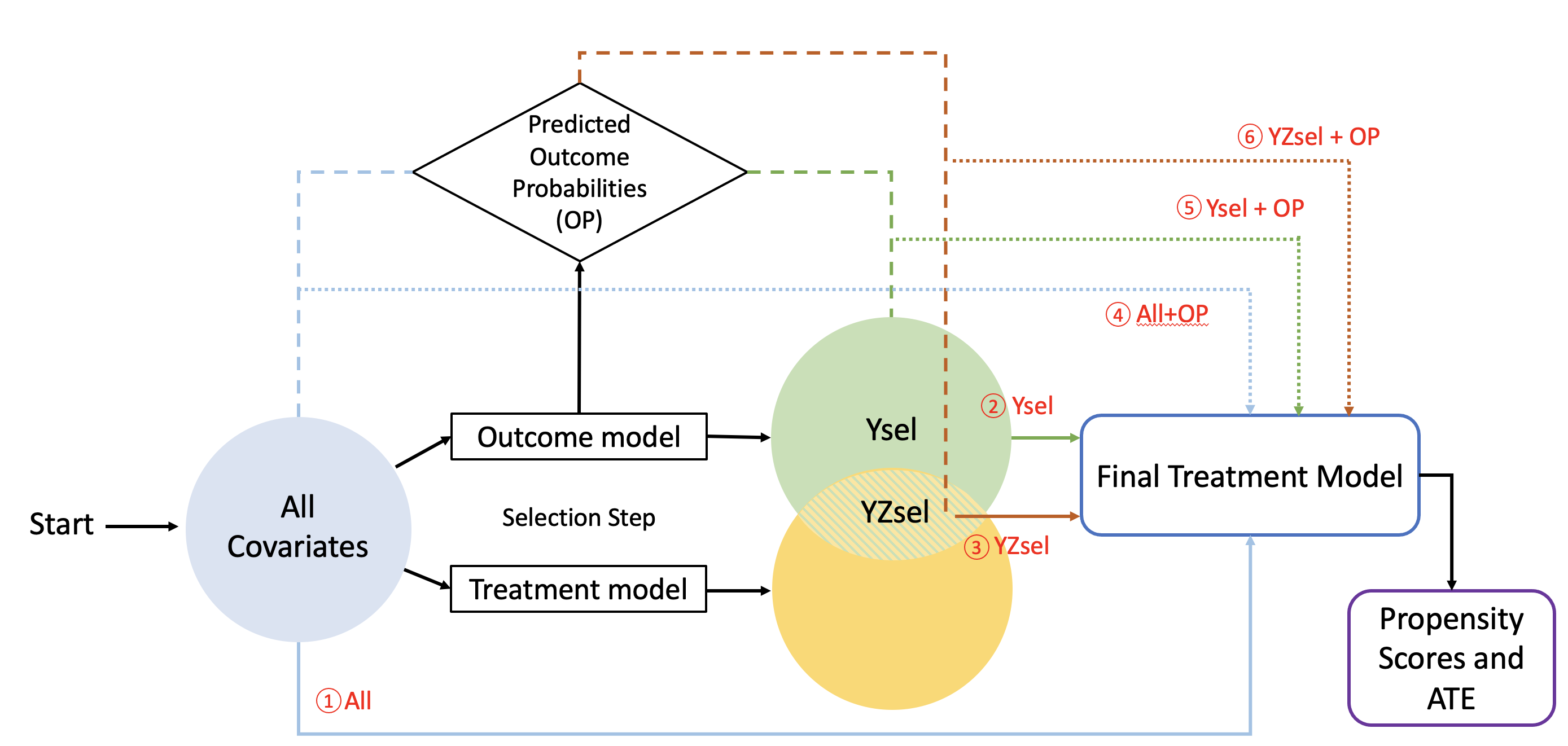}
    \caption{Flowchart for variable selection and propensity score estimation. Routes \textcircled{\scalebox{0.85}{1}}-\textcircled{\scalebox{0.85}{6}} correspond to different sets of input variables: \textcircled{\scalebox{0.85}{1}} All, \textcircled{\scalebox{0.85}{2}} Ysel, \textcircled{\scalebox{0.85}{3}} YZsel, \textcircled{\scalebox{0.85}{4}} OP+All, \textcircled{\scalebox{0.85}{5}} OP+Ysel, \textcircled{\scalebox{0.85}{6}} OP+YZsel.}
    \label{fig:Chap3_flowchart}
\end{figure}

When the dimension of the covariate vector is high, it tends to be infeasible to fit an unrestricted parametric model, such as a multinomial logistic regression model, for the treatment using all the available covariates. A practical problem for empirical researchers is to identify a subset of covariates to be conditioned on to control for confounding. Typically, in medical research, a list of important covariates will be suggested based on the evidence in the literature and/or expert opinion. However, as the number of available covariates increases, it becomes extremely difficult for human experts to check manually which variables are potential confounders. Alternatively, one can turn to data-driven variable selection approaches, such as LASSO \citep{tibshirani1996}, which automatically select important variables for treatment predictions from all the available covariates. Figure~\ref{fig:Chap3_flowchart} displays a flowchart of several possible routes that can be followed to identify the set of covariates to be included in the final treatment model for estimating the propensity scores. A commonly chosen route is to apply shrinkage methods directly to the original reservoir of covariates (route \textcircled{\scalebox{0.85}{1}} in Figure~\ref{fig:Chap3_flowchart}), and we label the set of covariates \verb|All|. In this case, variable selection and propensity score estimation are conducted simultaneously in a single step. However, as we noted later in our simulation, following route \textcircled{\scalebox{0.85}{1}} can result in substantially biased effect estimates. For sparse high-dimensional data, a large value of tuning parameter is necessary to select a parsimonious model. However, large penalties at the same time increase the shrinkage of non-zero components, leading to biased estimation \citep{meinshausen2007}. Therefore, reducing the number of spurious covariates entering the final treatment model through variable selection may help improve the performance of the propensity score estimation methods. In addition, such traditional variable selection approaches targeting $\boldsymbol{X}_\mathcal{C}\cup\boldsymbol{X}_\mathcal{Z}$ for the treatment only are sub-optimal. It has been shown that the use of $\boldsymbol{X}_\mathcal{Z}$ for propensity score modeling may inflate the variance of the estimated ATE \citep{brookhart2006, shortreed2017}. On the other hand, including $\boldsymbol{X}_\mathcal{Y}$ predictive of the outcome in the propensity score model can improve the precision of the ATE estimates \citep{brookhart2006, shortreed2017}. Therefore, an optimal propensity model for estimating ATE should include $\boldsymbol{X}_\mathcal{C}$ and $\boldsymbol{X}_\mathcal{Y}$ simultaneously while excluding $\boldsymbol{X}_\mathcal{Z}$.

\subsection{Using Outcome Model for Variable Selection}

To identify $\boldsymbol{X}_\mathcal{C}$, $\boldsymbol{X}_\mathcal{Y}$ and $\boldsymbol{X}_\mathcal{Z}$, we apply a pre-selection procedure to the original set of covariates (\verb|All|), where a regularized model (in this case we choose LASSO) is fitted separately for both the outcome and the treatment. Specifically, the model for a binary outcome is specified as
\begin{align}
    \text{logit}P(Y_i=1|\boldsymbol{X}_i)=\boldsymbol{X}_i^T\boldsymbol{\theta},
    \label{eq:Chap3_outcome_model}
\end{align}
the coefficients of which are estimated based on the LASSO penalty
\begin{align*}
    \hat{\boldsymbol{\theta}} = \argmin_{\boldsymbol{\theta}}\sum_{i=1}^n\left[\{-Y_i\boldsymbol{X}_i^T\boldsymbol{\theta}+\ln(1+e^{\boldsymbol{X}_i^T\boldsymbol{\theta}})\} + \lambda\sum_{k=1}^p |\theta_k|\right], 
\end{align*} 
where the tuning parameter $\lambda$ is chosen using cross-validation. Here we exclude the treatment variable $Z$ from the outcome model, and only consider the association of $\boldsymbol{X}$ with $Y$ across all treatment groups. For the treatment assignment mechanism, we assume a multinomial logistic regression model,
\begin{align}
P(Z_i=j|\boldsymbol{X}_i) = \frac{\exp(\boldsymbol{X}_i^T\boldsymbol{\psi}_j)}{\sum_{l=1}^J\exp(\boldsymbol{X}_i^T\boldsymbol{\psi}_l)}, 
\label{eq:Chap3_trt_model}
\end{align}
which is fitted by minimizing the negative penalized log-likelihood
\begin{align*}
    N^{-1}\sum_{i=1}^N\log P(Z_i=j|\boldsymbol{X}_i)
    + \lambda \sum_{k=1}^p \|\boldsymbol{\psi}_{\cdot k} \|_2 , 
\end{align*}
where $\boldsymbol{\psi}_{\cdot k}$ is a $J$-dimensional vector containing the $k$-th coefficient for all $J$ treatment groups \citep{friedman2010}. This group LASSO penalty function ensures that each variable in $\boldsymbol{X}_i$ is selected or excluded for all $J$ levels, as opposed to each level having its own set of selected variables. We use \verb|Ysel| to label the set of variables selected by Model \eqref{eq:Chap3_outcome_model} and \verb|YZsel| to label the intersection of the two sets of variables selected by Model \eqref{eq:Chap3_outcome_model} and Model \eqref{eq:Chap3_trt_model}. In an ideal case, \verb|YZsel| is identical to $\boldsymbol{X}_\mathcal{C}$ and \verb|Ysel|\textbackslash \verb|YZsel| is identical to $\boldsymbol{X}_\mathcal{Y}$. Theoretically, adjusting for $\boldsymbol{X}_\mathcal{C}$ alone in the treatment model is sufficient to remove the confounding bias. Route \textcircled{\scalebox{0.85}{3}} in Figure \ref{fig:Chap3_flowchart} corresponds to the case where \verb|YZsel| is used as predictors for propensity scores estimation.

As we mentioned previously, the ideal propensity score model adjusts for $\boldsymbol{X}_\mathcal{C}\cup\boldsymbol{X}_\mathcal{Y}$ while excluding $\boldsymbol{X}_\mathcal{Z}$. The OAL method proposed by \citet{shortreed2017} intends to achieve this by fitting an adaptive LASSO for the treatment \citep{zou2006}, where the adaptive weights are computed based on the coefficients from the outcome model, with smaller coefficients corresponding to larger weights. OAL discourages $\boldsymbol{X}_\mathcal{Z}$ from being selected and encourages the inclusion of $\boldsymbol{X}_\mathcal{Y}$ by imposing heavier penalties on $\boldsymbol{X}_\mathcal{Z}$ than $\boldsymbol{X}_\mathcal{Y}$. There are two limitations of the work of \citet{shortreed2017}. The first is that the coefficients of the outcome model were estimated using unpenalized linear regression, which may be fitted for a continuous outcome in the context where the ratio of $p$ to $n$ is relatively large. However, for a binary outcome as considered in our study, it tends to be more difficult for a standard logistic regression to converge. We extend their method by considering a LASSO-fitted outcome model to compute the adaptive weights, in which case the coefficients of covariates not predictive of the outcome can be zero, and therefore the covariates with zero coefficients will be excluded from the treatment model. Another limitation is that the idea of OAL cannot be conveniently extended to machine learning methods that do not rely on regularization.

OAL incorporates the outcome-covariate associations as adaptive weights when selecting variables for propensity score estimation. An alternative is to directly include the covariates predictive of the outcome in the treatment model, namely, \verb|Ysel|. The treatment can then be modeled as a function of \verb|Ysel| (route \textcircled{\scalebox{0.85}{2}} in Figure~\ref{fig:Chap3_flowchart}).

One possible problem of using \verb|Ysel| as input for the final treatment model is that the dimensionality of \verb|Ysel| can still remain relatively high for sample size $n$. To improve propensity score modeling by utilizing the variables predictive of the outcome only while maintaining a reasonable dimensionality, we propose to incorporate the information contained in $\boldsymbol{X}_\mathcal{Y}$ into the propensity score estimation process in the form of predicted probabilities of the outcome, Outcome Probability (OP) for short, as opposed to directly including those variables in their original form. We estimate OP for each subject by $\hat{p}_i=\exp(\boldsymbol{X}_i^T\hat{\boldsymbol{\theta}})/\{1+\exp(\boldsymbol{X}_i^T\hat{\boldsymbol{\theta}})\}$, and the logit scale of $\hat{p}_i$ is denoted $\hat{p}_i^\ast=\log\{\hat{p}_i/(1-\hat{p}_i)\}$. OP summarizes the information about the outcome-covariate relationship and reduces the dimensionality of the outcome predictors to a one-dimensional vector. After obtaining \verb|YZsel| which targets $\boldsymbol{X}_\mathcal{C}$, we add OP back to the set of input variables to capture the information in $\boldsymbol{X}_\mathcal{Y}$ (route \textcircled{\scalebox{0.85}{6}} in Figure \ref{fig:Chap3_flowchart}, and we denote the combination \verb|OP+YZsel|). In addition to \verb|YZsel|, as was noted later in the simulation studies, OP can also be combined with all the available covariates and \verb|Ysel| as input for the final treatment model in Figure~\ref{fig:Chap3_flowchart} to improve the effect estimates. We denote the combinations \verb|OP+All| and \verb|OP+Ysel|, which correspond to route \textcircled{\scalebox{0.85}{4}} and route \textcircled{\scalebox{0.85}{5}} in Figure~\ref{fig:Chap3_flowchart}, respectively. We discuss how OP can be used as predictors to estimate the propensity scores for different treatment modeling techniques in Section~\ref{s:Chap3_choices}, where we outline several possible choices for the final treatment model.

In summary, both \verb|Ysel| and \verb|YZsel| use the outcome model for selecting covariates to account for confounding. Variable selection and propensity score estimation are conducted separately in two steps for methods based on \verb|Ysel|, \verb|YZsel|, \verb|OP+Ysel|, and \verb|OP+YZsel|.

\section{Possible Choices for the Final Treatment Model for Propensity Score Estimation}
\label{s:Chap3_choices}

We consider five different methods (Table~\ref{tab:Chap3_methods_choices}) for estimating the propensity score given a set of candidate predictors. We leave out the OP for now and only consider the alternatives for the final treatment model for routes \textcircled{\scalebox{0.85}{1}}-\textcircled{\scalebox{0.85}{3}} in Figure~\ref{fig:Chap3_flowchart}. The first is a multinomial logistic regression model (LOGIS), specified as
\begin{align*}
    P(Z_i=j|\tilde{\boldsymbol{X}}_i) = \frac{\exp(\tilde{\boldsymbol{X}}_i^T\boldsymbol{\alpha}_j)}{\sum_{l=1}^J\exp(\tilde{\boldsymbol{X}}_i^T\boldsymbol{\alpha}_l)},
\end{align*}
where $\tilde{\boldsymbol{X}}_i$ is the set of (possibly selected) variables entering the final treatment model. When the dimension of $\tilde{\boldsymbol{X}}_i$ is sufficiently small, the coefficients can be estimated using the maximum likelihood estimator. In the case where dimension reduction is needed, the estimates $\hat{\boldsymbol{\alpha}}$ can be obtained by minimizing the negative log-likelihood with the group LASSO penalty
\begin{align*}
    N^{-1}\sum_{i=1}^N\log P(Z_i=j|\tilde{\boldsymbol{X}}_i)+\lambda \sum_{k=1}^p \|\boldsymbol{\alpha}_{\cdot k} \|_2,
\end{align*}
which indicates that each covariate in $\tilde{\boldsymbol{X}}_i$ is associated with all or none of the $J$ levels.

The second is the classification and regression tree (CART) method which performs recursive binary splitting on the feature space in a top-down fashion. At each split, CART agnostically searches for a variable $X$ and a cutpoint such that the response values in each of the resulting nodes lead to the greatest homogeneity \citep{breiman1984}. In that sense, CART intrinsically conducts variable selection while growing the tree, as variables that are not predictive of the treatment are less likely to be chosen at each split. We used the Gini index, a measure of the total variance across the $J$ classes, as the metrics for node splitting. Small value of Gini index indicates that observations in this node are predominated by a single class. CART tends to overfit the data. To address the overfitting issue, the common strategy is to first grow a large tree and prune it back in order to retain only part of the tree, as simpler trees tend to be less sensitive to the noises in the data. This method is referred to as pruned CART.

The single tree implementation of both CART and pruned CART, sometimes known as weak learners, may give poor predictions on their own. The ensemble methods, which combine multiple weak learners into one predictive model, have been developed to enhance the predictions. One example is the bootstrap aggregation of the CART algorithm (bagged CART). The bootstrap step of bagged CART involves randomly drawing $n$ observations (i.e., the same size as the original sample) with replacement from the original sample and fitting a CART separately for each bootstrap replicate. The bagging estimates of the probability of subjects being assigned to each class are obtained by averaging the predicted class probabilities from each of the single trees. Another popular ensemble method is the random forests. Similar to bagged CART, random forests build trees based on bootstrap samples of the original observations. What is different from bagged CART is that random forests only considers a random sample of $m$ predictors ($m<p$) at each split, and typically $m\approx\sqrt{p}$ is chosen for classification problems in practice \citep{james2021}. In our simulation studies and data example, we choose to grow a relatively large number of trees in order to stabilize the out-of-bag error rate.

\subsection{Implementation of Incorporating the Predicted Outcome Probabilities}

The methods listed in Table~\ref{tab:Chap3_methods_choices} can be applied to \verb|All|, \verb|Ysel|, and \verb|YZsel| for estimating the propensity scores, which are then used to compute the ATE. The OP is employed for propensity score estimation in different ways for different final treatment models. For the LOGIS method, the propensity scores are estimated by regressing the treatment variable on the union of the input covariates (\verb|All|, \verb|Ysel|, or \verb|YZsel|) and $\hat{p}_i^\ast$. When the regression model is regularized, no penalty is imposed on $\hat{p}_i^\ast$. In this way the outcome information is guaranteed to be utilized in the propensity score model.

For the tree-based machine learning methods such as CART, there is no straightforward way to force the OP into the tree growing process, where variable selection is intrinsically conducted. We instead conduct a two-step estimation procedure. We first obtain the set of propensity scores $\hat{\boldsymbol{\pi}}(\tilde{\boldsymbol{X}}_i)=\{\hat{\pi}_1(\tilde{\boldsymbol{X}}_i),\cdots,\hat{\pi}_{J-1}(\tilde{\boldsymbol{X}}_i)\}^T$ using the tree-based methods.  Next, we fit a logistic regression model for the treatment as a function of $\hat{p}_i^\ast$ and the propensity scores obtained in route \textcircled{\scalebox{0.85}{1}}, \textcircled{\scalebox{0.85}{2}}, or \textcircled{\scalebox{0.85}{3}}:

\begin{align}
    \log\frac{P(Z_i=j|\tilde{\boldsymbol{X}}_i)}{P(Z_i=J|\tilde{\boldsymbol{X}}_i)}=\phi_{0j} + \hat{\boldsymbol{\pi}}(\tilde{\boldsymbol{X}}_i)^T\boldsymbol{\eta}_j+\phi_j\hat{p}_i^\ast.
    \label{eq:Chap3_OA_trt_model}
\end{align}
The coefficients $(\boldsymbol{\phi}_0,\boldsymbol{\phi}, \boldsymbol{\eta}_1^T,\cdots,\boldsymbol{\eta}_{J-1})^T$ can be estimated using maximum likelihood. The final propensity scores that take OP into account are then obtained by calculating the predicted probabilities from model (\ref{eq:Chap3_OA_trt_model}).

For \verb|OP+Ysel| (route \textcircled{\scalebox{0.85}{5}}) and \verb|OP+YZsel| (route \textcircled{\scalebox{0.85}{6}}), the associations between the covariates and the outcome are used twice in the entire estimation process, one for variable selection using the outcome model, and the other for propensity score estimation using the OP.

\begin{table}[h]
\resizebox{\textwidth}{!}{%
\begin{tabular}{lccc}
\hline
\multicolumn{1}{c}{\begin{tabular}[c]{@{}c@{}}Methods for Constructing\\ the Treatment Model\end{tabular}} & \begin{tabular}[c]{@{}c@{}}Simultaneous Variable \\ Selection and Estimation\end{tabular} & \begin{tabular}[c]{@{}c@{}}Ways to Incorporate OP\\ into the Estimation Process\end{tabular} & R package \\ \hline
\begin{tabular}[c]{@{}l@{}}Logistic regression\\ (LOGIS, standard or penalized)\end{tabular} & \begin{tabular}[c]{@{}c@{}}Yes for penalized \\ logistic regression\end{tabular} & \begin{tabular}[c]{@{}c@{}}Used as a regressor for logistic regression.\\ No penalty is imposed on OP \\ for penalized logistic regression.\end{tabular} & glmnet*, gcdnet \\ \hline
CART & Yes & \begin{tabular}[c]{@{}c@{}}A multinomial logistic regression for the treatment is fitted\\ as a function of estimated propensity scores and OP\end{tabular} & rpart* \\ \hline
Pruned CART & Yes & Same as above & rpart* \\ \hline
Bagged CART & Yes & Same as above & ipred* \\ \hline
Random forests & Yes & Same as above & randomForest*, ranger \\ \hline
\end{tabular}%
}
\caption{Possible choices for the final treatment model.\\
* R packages used to implement the methods in this paper.}
\label{tab:Chap3_methods_choices}
\end{table}

\section{Simulation Studies}
\label{s:Chap3_simulation}

\subsection{Implementation of Methods under Comparison}
The comparative methods were implemented in R with default parameters unless otherwise specified. The LASSO algorithm was implemented using the R package \textit{glmnet}. The tuning parameter $\lambda$ was determined using 10-fold cross-validation with the \verb+lambda.1se+ criterion for selecting variables in the pre-selection step, as the goal was to reduce the number of covariates entering the final treatment model. For LOGIS based on \verb|All|, which does not involve the pre-selection step, the \verb+lambda.min+ criterion was used. CART was implemented using the \textit{rpart} package with the complexity parameter (cp) being 0.001, which encourages a large and complex tree structure. For pruned CART, the ``cp'' that corresponded to the smallest 10-fold cross-validated error was used to determine the best pruned tree. Bagged CART was implemented using the \textit{ipred} package with 200 bootstrap replicates. Random forests were implemented using the \textit{randomForest} package with 1000 bootstrap replicates. The minimum size of terminal nodes (the \verb+nodesize+ parameter) was set to be 7 in order to make it consistent with the parameters used for CART. For bagged CART and random forests, propensity scores were estimated based on the out-of-bag predictions.

We also extended the OAL approach to three-treatment comparison. Following \citet{shortreed2017}, we considered a set of possible values for the tuning parameter $\lambda_n$, $\{n^{-20}, n^{-15}, n^{-10}, n^{-5}, n^{-3}, n^{-1},\\ n^{-0.75}, n^{-0.5}, n^{-0.25}, n^{0.25}, n^{0.49}\}$, and $\lambda_n$ was selected by minimizing a weighted absolute mean difference between treatment groups, a quantity that combines the weighted difference in covariates and the absolute values of the coefficients corresponding to the covariates in the outcome model. Since the adaptive weights for the covariates excluded by the outcome model got inflated to infinity, these covariates were not used to fit the adaptive LASSO for variable selection and propensity score estimation. In that sense, OAL in the high-dimensional setting is equivalent to a LASSO based on \verb|Ysel| with additional weights imposed on the covariates for variable selection.

\subsection{Simulation Setup}
For each simulated dataset, $J=3$ treatment groups were compared and $p=100$ covariates were considered, with $|\mathcal{C}|$ confounders, $|\mathcal{Z}|$ related to treatment only, $|\mathcal{Y}|$ related to outcome only, and $|\mathcal{S}|$ spurious predictors. Covariates $\boldsymbol{X}_i$ were generated as follows unless otherwise specified. Half of the covariates (rounded down) in $\boldsymbol{X}_\mathcal{C}$, $\boldsymbol{X}_\mathcal{Z}$, $\boldsymbol{X}_\mathcal{Y}$, and $\boldsymbol{X}_\mathcal{S}$ were generated from a binomial distribution with a probability of 0.3, and the other half were generated from a multivariate normal distribution with a $p'$-dimensional vector of means $\boldsymbol{0}_{p'}$ and a covariance matrix $\Sigma$, where $p'$ is the number of continuous covariates in each subset ($\boldsymbol{X}_\mathcal{C}$, $\boldsymbol{X}_\mathcal{Z}$, $\boldsymbol{X}_\mathcal{Y}$, or $\boldsymbol{X}_\mathcal{S}$) and $\Sigma$ is an identity matrix. 

We considered three different simulation settings. Our first setting assumes additivity and linearity for the treatment generating model,
\begin{align}
    Z_i\sim\text{Multinomial}\{\pi_1(\boldsymbol{V}_i), \pi_2(\boldsymbol{V}_i), \pi_3(\boldsymbol{V}_i)\},
    \label{eq:Chap3:trt_generating_model}
\end{align}
where 
$\pi_j(\boldsymbol{V}_i)=\exp(\boldsymbol{V}_i^T\boldsymbol{\alpha}_j)/\sum_{l=1}^3\exp(\boldsymbol{V}_i^T\boldsymbol{\alpha}_l)$ is the probability of receiving treatment $j$, with $\boldsymbol{V}_i=(1, \boldsymbol{X}_{\mathcal{C}}^T, \boldsymbol{X}_{\mathcal{Z}}^T)^T$. We assumed heterogeneous treatment effects on the outcome, and sampled the potential outcome $Y_i^{(j)}$ from a binomial distribution with probability 
\begin{align*}
    P\{Y_i^{(j)}=1|\boldsymbol{W}_i\}=\text{expit}\{\beta_{0j}+\boldsymbol{W}_i^T\boldsymbol{\beta}_j\},
\end{align*}
where $\boldsymbol{W}_i=(\boldsymbol{X}_\mathcal{C}^T, \boldsymbol{X}_\mathcal{Y}^T)^T$, $(\beta_{01}, \beta_{02}, \beta_{03})=(0, 0.6, 0.4)$, and $\boldsymbol{\beta}_j\propto(1,\cdots,1)^T$. We considered three levels of sparsity for the treatment and outcome models, with $|\mathcal{C}|=|\mathcal{Z}|=|\mathcal{Y}|=5$, $10$, and $20$ for the scenarios with sparse, moderately sparse, and dense models, respectively. As a result, the dimension of $\boldsymbol{\alpha}_j$ differed across the scenarios. The parameter $\boldsymbol{\alpha}=(\boldsymbol{\alpha}_1^T,\boldsymbol{\alpha}_2^T, \boldsymbol{\alpha}_3^T)^T$ was scaled such that $\|\boldsymbol{\alpha}\|_2=5$. Similarly, the signal strength of the outcome model was scaled such that $\|\boldsymbol{\beta}_1\|=3$, $\|\boldsymbol{\beta}_2\|=2$, and $\|\boldsymbol{\beta}_3\|=4$. The true values for $E\{Y^(1)\}$, $E\{Y^(2)\}$, and $E\{Y^(3)\}$ were 0.61, 0.71, and 0.68 for the `sparse' scenario, 0.69, 0.77, and 0.76 for the `moderately sparse' scenario, and 0.76, 0.83, and 0.83 for the `dense' scenario. To examine the performance as the sample size increases, we let the sample size $n$ be 500, 1000, and 2000 for the `sparse' scenario.

Our second setting assumed that models were `sparse' ($|\mathcal{C}|=|\mathcal{Z}|=|\mathcal{Y}|=5$) and considered a set of association equations for the treatment assignment that varied in degrees of nonlinearity and nonadditivity. In this case, $\boldsymbol{V}_i$ in (\ref{eq:Chap3:trt_generating_model}) may contain some transformation of covariates in $(\boldsymbol{X}_\mathcal{C}, \boldsymbol{X}_\mathcal{Y})$ and/or interaction effects. The structure of $\boldsymbol{V}_i$ are shown in Table A.1. Specifically, we considered scenarios with nonlinear main effects and no interactions (NL), linear main and interaction effects (L-L), nonlinear main effects and linear interactions (NL-L), and nonlinear main and interaction effects (NL-NL). All confounders were continuous in this setting, and the outcomes were generated using the same model as was used in the first setting except that $(\beta_{01}, \beta_{02}, \beta_{03})=(0, -0.6, 0.4)$.

Our third setting also assumed that the models were sparse and let censoring come into play. Instead of sampling binary outcomes directly from a binomial distribution, we first generated time to event $T_i$ from a logistic distribution with mean function
\begin{align*}
    \beta_{01}I(Z_i=1)+\beta_{02}I(Z_i=2)+\beta_{03}I(Z_i=3)+\boldsymbol{W}_i^T\boldsymbol{\beta}
\end{align*}
and scale parameter $s=6$, where $\boldsymbol{W}_i=(\boldsymbol{X}_\mathcal{C}^T, \boldsymbol{X}_\mathcal{Y}^T)^T$, $(\beta_{01}, \beta_{02}, \beta_{03})=(120,100,115)$ and $\boldsymbol{\beta}=(5,\cdots,5)^T$. Then we obtained the outcome such that $Y_i=I\{T_i<130\}$. We generated the censoring time $C$ using inverse transform sampling \citep{Bender2005} as a function of $\boldsymbol{U}_i=(1,\boldsymbol{X}_\mathcal{C})^T$. Specifically, we assumed a Cox proportional hazards model with the baseline hazard following a Weibull distribution,
\begin{align*}
    C_i^{(j)} = \{\lambda^{-1}\exp(\boldsymbol{U}_i^T\boldsymbol{\gamma}_j)^{-1}\log u\}^{1/\nu},
\end{align*}
with scale parameter $\lambda=0.01$ and shape parameter $\nu=7$, where $u$ was sampled from a $\text{Uniform}(0,1)$ distribution. Note that $\boldsymbol{U}_i$ was assumed to be known in our simulation settings, which resembles our data example where censoring is believed to only depend a low-dimensional set of covariates that can be identified by human experts. The proportion of subject not being observed for $d=130$ was around 22\%.

We considered the six routes displayed in Figure~\ref{fig:Chap3_flowchart} for propensity score estimation. We also present the results for three sets of covariates that are usually unknown in practice: confounders only ($\boldsymbol{X}_\mathcal{C}$), treatment predictors ($\boldsymbol{X}_\mathcal{C}\cup\boldsymbol{X}_\mathcal{Z}$), and outcome predictors ($\boldsymbol{X}_\mathcal{C}\cup\boldsymbol{X}_\mathcal{Y}$). These results were used to illustrate the impact of different groups of covariates on the treatment effect estimates. The ATE were estimated using the IPW method. 

\subsection{Construction of Confidence Intervals}
The confidence intervals (CI) were constructed using bootstrap standard errors based on 200 bootstrap replicates. For the settings without censoring, we considered two possible bootstrap procedures. The first applies variable selection to each bootstrap replicate and refits the models for the treatment and outcome using variables selected in each replicate, and we refer to it as usual bootstrap. The second ignores the variability due to selection for covariates. Instead, for LOGIS based on \verb|All| and \verb|OP+All| and all tree-based methods, OP and propensity scores are directly bootstrapped from the OP and propensity scores obtained in the original sample, without refitting the model. For LOGIS based on \verb|Ysel|, \verb|YZsel|, \verb|OP+Ysel|, and \verb|OP+YZsel|, propensity scores are obtained by refitting the final treatment model using variables selected in the original data set. As a result, the usual bootstrap is much more computationally intensive than the modified bootstrap. A trial simulation study under the scenario of linear sparse models for sample sizes of 500 (Table \ref{tab:Chap3_bootstrap_n500}) and 1000 (Table A.2) showed that usual bootstrap tended to overestimate the standard errors and produce overly conservative CIs for the tree-based methods. For example, most of the coverage rates for the methods that involved OP were above 98\%. Over-coverage was also observed for (unpenalized) LOGIS for usual bootstrap. On the other hand, LOGIS based on \verb|OP+All| had close to nominal coverage using usual bootstrap at the expense of high computational burden, and slight over-coverage using modified bootstrap. For the tree-based methods, standard errors based on modified bootstrap were close to their corresponding Monte Carlo standard deviation. In this case, modified bootstrap remedied the overestimation of the usual bootstrap by dropping the variability due to variable selection and estimated the true variability well across all methods. Therefore, we proceed with the modified bootstrap technique for our simulation studies, which directly samples the estimated propensity scores and OP from the original simulated data set for each bootstrap replicate. For the third setting where censoring existed, we again used modified bootstrap, but with the censoring model refitted for each bootstrap sample. Metrics that were used to compare the various propensity score estimation methods included bias, Monte Carlo standard deviations, standard errors, root mean squared error (RMSE), and coverage rate of 95\% CIs. True values were determined using $5\times10^5$ replicates.

\begin{table}[h]
\resizebox{\textwidth}{!}{%
\begin{tabular}{rcccccccccccccccccc}
\toprule
\multicolumn{1}{l}{}                     & \multicolumn{3}{c}{\begin{tabular}[c]{@{}c@{}}Bias\\      $\times$1000\end{tabular}} & \multicolumn{3}{c}{\begin{tabular}[c]{@{}c@{}}Empirical SD\\      $\times$1000\end{tabular}} & \multicolumn{3}{c}{\begin{tabular}[c]{@{}c@{}}SE (usual)\\      $\times$1000\end{tabular}} & \multicolumn{3}{c}{\begin{tabular}[c]{@{}c@{}}Coverage (\%)\\  (usual)    \end{tabular}} & \multicolumn{3}{c}{\begin{tabular}[c]{@{}c@{}}SE (modified)\\      $\times$1000\end{tabular}} & \multicolumn{3}{c}{\begin{tabular}[c]{@{}c@{}}Coverage (\%)\\ (modified) \end{tabular}} \\ \cmidrule(lr){2-4} \cmidrule(lr){5-7} \cmidrule(lr){8-10} \cmidrule(lr){11-13} \cmidrule(lr){14-16} \cmidrule(lr){17-19} 
\multicolumn{1}{l}{Estimators}            & 1 vs 2                   & 1 vs 3                  & 2 vs 3                  & 1 vs 2                     & 1 vs 3                     & 2 vs 3                     & 1 vs 2                     & 1 vs 3                    & 2 vs 3                    & 1 vs 2                      & 1 vs 3                     & 2 vs 3                     & 1 vs 2                      & 1 vs 3                     & 2 vs 3                     & 1 vs 2                      & 1 vs 3                     & 2 vs 3  
\\ \toprule
\multicolumn{1}{l}{\textbf{OAL}}         & 101                      & 77                      & -25                     & 54                         & 48                         & 51                         & 57                         & 51                        & 53                        & 96.1                        & 96.0                       & 95.7                       & 55                          & 50                         & 52                         & 94.1                        & 95.6                       & 94.2                       \\
\multicolumn{1}{l}{\textbf{LOGIS}}       &                          &                         &                         &                            &                            &                            &                            &                           &                           &                             &                            &                            &                             &                            &                            &                             &                            &                            \\
All                                      & 96                       & 75                      & -21                     & 52                         & 48                         & 49                         & 56                         & 54                        & 53                        & 59.4                        & 73.2                       & 93.4                       & 55                          & 55                         & 52                         & 59.8                        & 75.9                       & 93.3                       \\
Ysel                                     & -6                       & -2                      & 5                       & 56                         & 49                         & 55                         & 94                         & 82                        & 85                        & 100.0                       & 99.7                       & 99.9                       & 58                          & 51                         & 55                         & 95.4                        & 96.0                       & 94.8                       \\
YZsel                                    & -1                       & 3                       & 4                       & 57                         & 50                         & 55                         & 80                         & 70                        & 74                        & 99.7                        & 99.1                       & 99.4                       & 58                          & 53                         & 55                         & 95.0                        & 95.5                       & 95.0                       \\
OP+All                                   & 15                       & 15                      & 0                       & 50                         & 45                         & 49                         & 57                         & 52                        & 54                        & 95.8                        & 96.9                       & 95.5                       & 58                          & 56                         & 56                         & 96.3                        & 97.8                       & 96.9                       \\
OP+Ysel                                  & -6                       & -2                      & 5                       & 56                         & 49                         & 55                         & 94                         & 82                        & 85                        & 100.0                       & 99.7                       & 99.9                       & 58                          & 51                         & 55                         & 95.4                        & 96.0                       & 94.8                       \\
OP+YZsel                                 & -5                       & -1                      & 5                       & 54                         & 47                         & 54                         & 80                         & 69                        & 73                        & 99.8                        & 99.4                       & 99.3                       & 55                          & 49                         & 53                         & 95.2                        & 96.0                       & 94.3                       \\
\multicolumn{1}{l}{\textbf{CART}}        &                          &                         &                         &                            &                            &                            &                            &                           &                           &                             &                            &                            &                             &                            &                            &                             &                            &                            \\
All                                      & 125                      & 95                      & -30                     & 79                         & 78                         & 74                         & 95                         & 96                        & 89                        & 78.1                        & 87.4                       & 97.0                       & 78                          & 80                         & 74                         & 63.2                        & 75.9                       & 92.0                       \\
Ysel                                     & 74                       & 57                      & -17                     & 68                         & 68                         & 66                         & 87                         & 86                        & 82                        & 90.9                        & 94.4                       & 97.5                       & 72                          & 72                         & 69                         & 81.6                        & 88.1                       & 94.0                       \\
YZsel                                    & 59                       & 45                      & -14                     & 66                         & 62                         & 62                         & 83                         & 83                        & 79                        & 93.1                        & 96.4                       & 99.0                       & 70                          & 68                         & 67                         & 86.2                        & 90.8                       & 94.8                       \\
OP+All                                   & 10                       & 10                      & 0                       & 77                         & 73                         & 78                         & 101                        & 93                        & 102                       & 98.8                        & 98.6                       & 98.9                       & 79                          & 74                         & 79                         & 93.4                        & 93.8                       & 94.2                       \\
OP+Ysel                                  & 14                       & 12                      & -3                      & 63                         & 61                         & 64                         & 87                         & 81                        & 88                        & 99.4                        & 99.0                       & 98.7                       & 66                          & 63                         & 66                         & 94.2                        & 94.4                       & 94.8                       \\
OP+YZsel                                 & 14                       & 10                      & -4                      & 60                         & 56                         & 61                         & 82                         & 77                        & 84                        & 99.3                        & 99.5                       & 99.4                       & 63                          & 59                         & 62                         & 94.3                        & 94.7                       & 94.8                       \\
\multicolumn{1}{l}{\textbf{Pruned CART}} &                          &                         &                         &                            &                            &                            &                            &                           &                           &                             &                            &                            &                             &                            &                            &                             &                            &                            \\
All                                      & 134                      & 102                     & -32                     & 63                         & 61                         & 55                         & 92                         & 92                        & 86                        & 75.3                        & 90.4                       & 99.1                       & 61                          & 61                         & 56                         & 40.1                        & 60.4                       & 90.4                       \\
Ysel                                     & 97                       & 74                      & -23                     & 61                         & 59                         & 54                         & 85                         & 84                        & 79                        & 87.0                        & 93.2                       & 99.0                       & 60                          & 60                         & 56                         & 61.0                        & 75.5                       & 92.4                       \\
YZsel                                    & 81                       & 63                      & -19                     & 64                         & 60                         & 54                         & 82                         & 81                        & 76                        & 87.2                        & 93.5                       & 99.3                       & 61                          & 60                         & 57                         & 68.6                        & 80.7                       & 93.0                       \\
OP+All                                   & -3                       & 1                       & 4                       & 56                         & 51                         & 55                         & 95                         & 87                        & 96                        & 99.8                        & 99.5                       & 99.9                       & 58                          & 52                         & 57                         & 94.3                        & 94.5                       & 94.6                       \\
OP+Ysel                                  & 1                        & 4                       & 2                       & 52                         & 49                         & 54                         & 83                         & 77                        & 84                        & 100.0                       & 99.5                       & 99.7                       & 55                          & 50                         & 54                         & 94.2                        & 94.8                       & 94.6                       \\
OP+YZsel                                 & 2                        & 3                       & 1                       & 53                         & 48                         & 53                         & 78                         & 73                        & 80                        & 99.8                        & 99.6                       & 99.6                       & 55                          & 50                         & 54                         & 95.7                        & 94.9                       & 94.7                       \\
\multicolumn{1}{l}{\textbf{Bagged CART}} &                          &                         &                         &                            &                            &                            &                            &                           &                           &                             &                            &                            &                             &                            &                            &                             &                            &                            \\
All                                      & 120                      & 91                      & -29                     & 54                         & 51                         & 49                         & 45                         & 45                        & 42                        & 28.9                        & 46.9                       & 85.3                       & 56                          & 57                         & 53                         & 44.1                        & 65.5                       & 91.2                       \\
Ysel                                     & 17                       & 17                      & 0                       & 67                         & 61                         & 64                         & 54                         & 51                        & 49                        & 87.3                        & 88.7                       & 86.2                       & 71                          & 66                         & 67                         & 93.4                        & 95.6                       & 95.4                       \\
YZsel                                    & -35                      & -26                     & 8                       & 90                         & 80                         & 85                         & 63                         & 59                        & 57                        & 80.0                        & 82.1                       & 80.9                       & 86                          & 79                         & 81                         & 91.6                        & 92.5                       & 94.8                       \\
OP+All                                   & -5                       & 1                       & 6                       & 51                         & 46                         & 50                         & 86                         & 84                        & 90                        & 100.0                       & 99.9                       & 99.9                       & 52                          & 47                         & 51                         & 94.2                        & 95.4                       & 94.4                       \\
OP+Ysel                                  & -7                       & 0                       & 7                       & 47                         & 44                         & 48                         & 80                         & 80                        & 85                        & 100.0                       & 99.9                       & 100.0                      & 50                          & 45                         & 50                         & 94.6                        & 95.2                       & 95.0                       \\
OP+YZsel                                 & -7                       & -1                      & 6                       & 47                         & 44                         & 49                         & 77                         & 77                        & 82                        & 100.0                       & 99.8                       & 99.9                       & 50                          & 46                         & 50                         & 94.6                        & 95.0                       & 94.6                       \\
\multicolumn{2}{l}{\textbf{Random   Forests}}                       &                         &                         &                            &                            &                            &                            &                           &                           &                             &                            &                            &                             &                            &                            &                             &                            &                            \\
All                                      & 135                      & 103                     & -32                     & 50                         & 48                         & 47                         & 40                         & 40                        & 38                        & 13.4                        & 30.9                       & 80.4                       & 52                          & 53                         & 50                         & 27.1                        & 51.2                       & 90.0                       \\
Ysel                                     & 53                       & 43                      & -10                     & 56                         & 50                         & 51                         & 41                         & 41                        & 38                        & 68.4                        & 76.2                       & 84.7                       & 59                          & 57                         & 57                         & 84.9                        & 91.2                       & 95.9                       \\
YZsel                                    & -36                      & -22                     & 14                      & 91                         & 72                         & 78                         & 46                         & 44                        & 41                        & 66.8                        & 75.7                       & 72.0                       & 79                          & 70                         & 75                         & 91.0                        & 91.2                       & 94.5                       \\
OP+All                                   & -5                       & 1                       & 6                       & 53                         & 46                         & 51                         & 61                         & 56                        & 59                        & 97.5                        & 98.4                       & 97.2                       & 53                          & 48                         & 52                         & 94.8                        & 94.9                       & 94.0                       \\
OP+Ysel                                  & -9                       & -1                      & 8                       & 48                         & 44                         & 49                         & 70                         & 66                        & 71                        & 99.9                        & 99.5                       & 99.4                       & 51                          & 46                         & 50                         & 94.5                        & 94.8                       & 94.4                       \\
OP+YZsel                                 & -8                       & -1                      & 7                       & 48                         & 45                         & 49                         & 79                         & 76                        & 82                        & 99.9                        & 99.7                       & 99.9                       & 51                          & 46                         & 50                         & 94.8                        & 95.0                       & 94.7      \\ \toprule                 
\end{tabular}
}
\caption[Standard errors (SE) and coverage of 95\% confidence intervals estimated by usual bootstrap and modified bootstrap for sample size of 500.]{Standard errors (SE) and coverage of 95\% confidence intervals estimated by usual bootstrap and modified bootstrap for sample size of 500. The scenario with sparse treatment models was considered. Results were obtained based on 1000 simulated datasets. For each dataset, 200 bootstrap samples were generated.}
\label{tab:Chap3_bootstrap_n500}
\end{table}

\subsection{Simulation Results}
We present the box plots of bias for the IPW estimates across the simulation settings in Figures \ref{fig:Chap3_bias_sparsity_OP}-\ref{fig:Chap3_bias_nonlinear_OP}. The numerical results for all evaluation metrics are reported in the Supplementary Materials.

\subsubsection{Bias}
\begin{figure}[p]
    \centering
    \includegraphics[width=\textwidth]{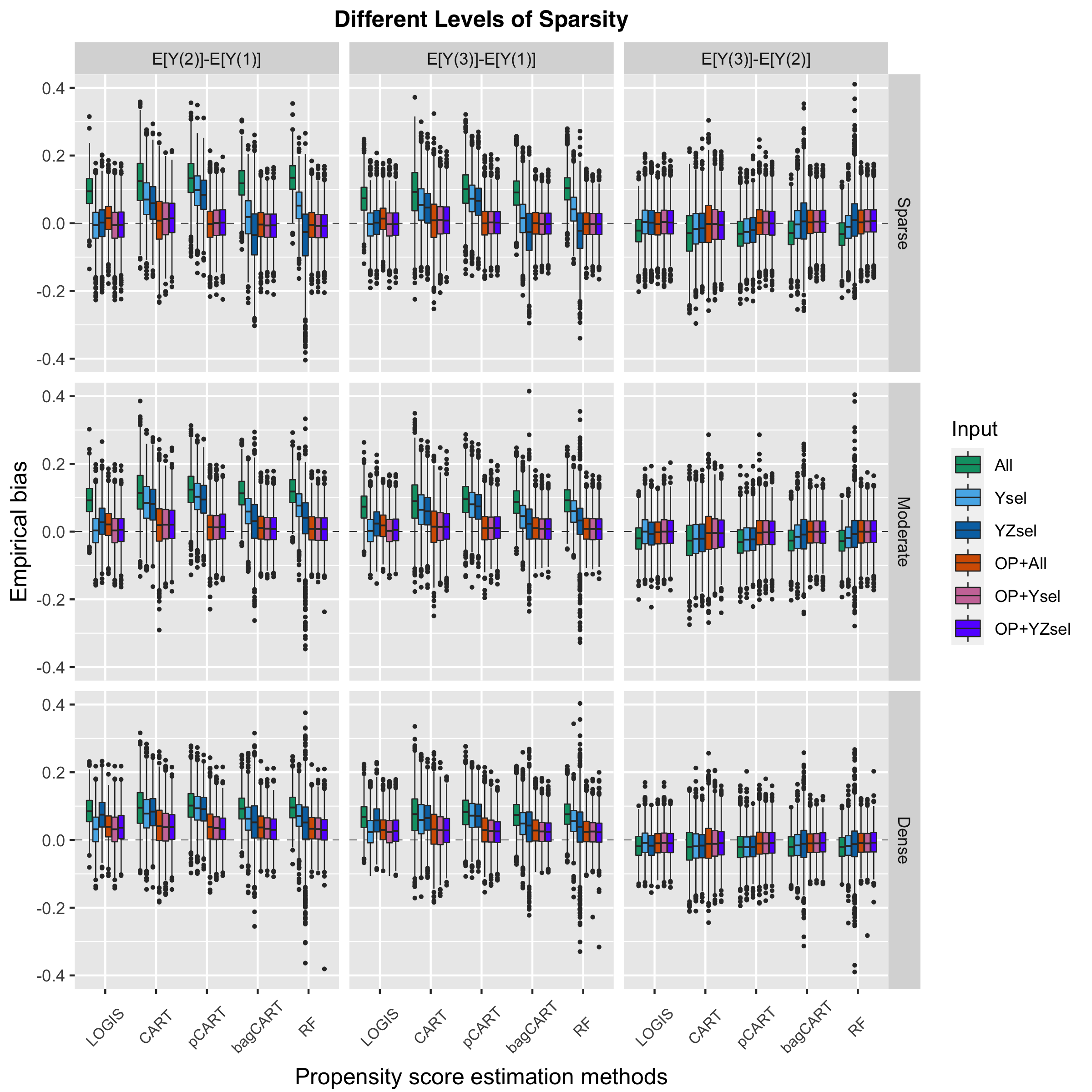}
    \caption[Box plots of empirical bias for 2000 inverse probability weighted estimates for the ATE under scenarios with different levels of sparsity.]{Box plots of empirical bias for 2000 inverse probability weighted estimates for the ATE under scenarios with different levels of sparsity. The rows represent scenarios and columns represent treatment pairs. Each simulated dataset contained 500 samples.}
    \label{fig:Chap3_bias_sparsity_OP}
\end{figure}

The numerical results for the setting of linear treatment models are presented in Tables A.3-A.9 in the Supplementary Materials. LOGIS that adjusted for confounders ($\boldsymbol{X}_\mathcal{C}$), treatment predictors ($\boldsymbol{X}_\mathcal{C}\cup\boldsymbol{X}_\mathcal{Z}$), and outcome predictors ($\boldsymbol{X}_\mathcal{C}\cup\boldsymbol{X}_\mathcal{Y}$) had close to zero empirical bias as expected (e.g., Table A.3). For the CART family methods, using confounders only generally yielded smaller bias than using treatment predictors, outcome predictors, or \verb+All+. Methods based on pre-selected predictors (\verb+Ysel+ and \verb+YZsel+) tended to yield smaller empirical bias than methods using all the available predictors as input (Figure~\ref{fig:Chap3_bias_sparsity_OP}), which indicates that excluding noise variables before fitting the final treatment model can help remove the bias, though the absolute bias was still greater than zero for the tree-based methods. One exception was that random forests based on \verb+YZsel+ resulted in substantial bias in the `sparse' scenario, possibly because there were too few true predictors at each split for random forests to choose from. The OAL method had similar performance in terms of bias to LOGIS based on \verb|OP+Ysel| and \verb|OP+YZsel| in the `sparse' scenario (Tables A.3-A.5), while the latter outperformed the former in the `moderately sparse' (Table A.6) and `dense' scenarios (Tables A.8 and A.9).

\begin{figure}[p]
    \centering
    \includegraphics[width=\textwidth]{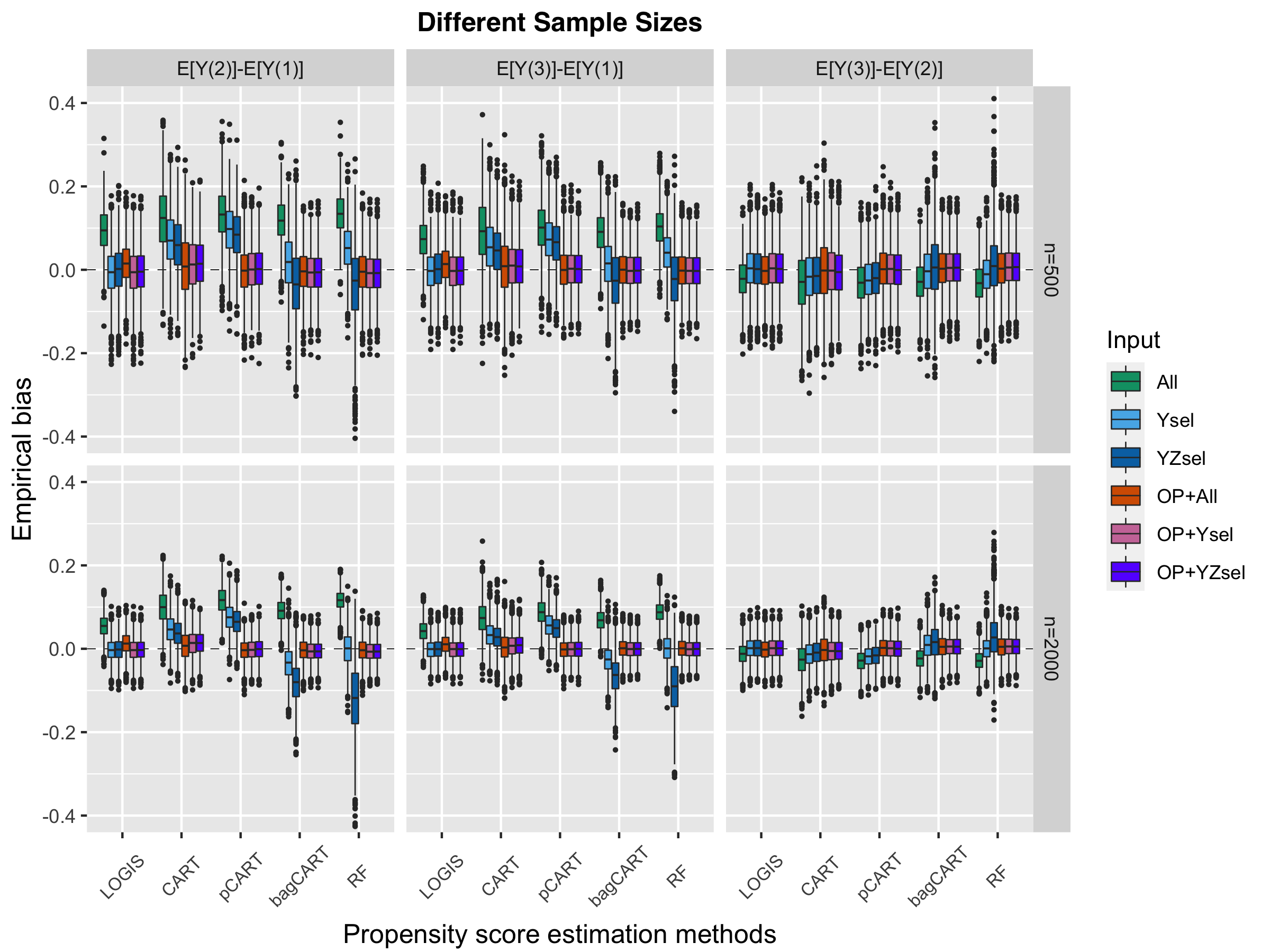}
    \caption[Box plots of empirical bias for 2000 inverse probability weighted estimates for the ATE for different sample sizes.]{Box plots of empirical bias for 2000 inverse probability weighted estimates for the ATE for different sample sizes. The scenario with sparse treatment and outcome models was considered. The rows represent sample sizes and columns represent treatment pairs.}
    \label{fig:Chap3_bias_sample_size_OP}
\end{figure}

When the outcome information was not taken into account, LOGIS in general produced less biased estimates than the tree-based methods for each of the routes \textcircled{\scalebox{0.85}{1}}-\textcircled{\scalebox{0.85}{3}} in Figure~\ref{fig:Chap3_flowchart}. Bias for \verb+All+, \verb+Ysel+, and \verb+YZsel+ were getting closer as the model became `denser' (Figure \ref{fig:Chap3_bias_sparsity_OP}). As sample size increased from 500 to 2000 for the `sparse' scenario (Figure \ref{fig:Chap3_bias_sample_size_OP}), the nonzero empirical bias persisted across the methods considered, which illustrates the slow convergence rate of LASSO. The inclusion of OP greatly reduced the bias of the estimates in the case where methods based on \verb|All|, \verb|Ysel|, or \verb|YZsel| resulted in larger than zero absolute bias across the simulation settings, which highlights the robustness of the treatment effect estimator provided by incorporating OP into the estimation process.

With nonlinearity and nonadditivity in the treatment model, the performance of LOGIS was not inferior to the tree-based methods in terms of bias (Tables A.10-A.17), possibly because in this case, LOGIS approximated nonlinear functions reasonably well. The good approximation of linear methods to nonlinear functions was also observed in \citet{tu2019a}, where multivariable linear regression yielded smaller bias than bagged CART and random forests in some cases with nonlinear and nonadditive associations in the treatment models. 

When censoring existed, LOGIS based on \verb|Ysel|, \verb|YZsel|, \verb|OP+Ysel|, and \verb|OP+YZsel| produced much lower bias than their CART family counterparts. One exception was \verb|YZsel|-based bagged CART, which had close to zero bias (Table A.18).

\subsubsection{Statistical Efficiency and RMSE}
In general, estimates produced by routes \textcircled{\scalebox{0.85}{4}}, \textcircled{\scalebox{0.85}{5}}, and \textcircled{\scalebox{0.85}{6}} had smaller variability than those resulting from routes \textcircled{\scalebox{0.85}{1}}, \textcircled{\scalebox{0.85}{2}} and \textcircled{\scalebox{0.85}{3}}, respectively, for each method across all the settings, which illustrates the advantage of leveraging the information about the outcome model in terms of statistical efficiency.

Methods based on \verb|OP+All|, \verb|OP+Ysel|, and \verb|OP+YZsel| had smaller variabilities and RMSE of the effect estimates across the simulation settings compared to methods based on \verb|All|, \verb|Ysel|, and \verb|YZsel|, respectively (Figures B.1-B.3). For example, for the scenario with moderately sparse models, the percentage of reduction in RMSE ranged from 1.6\% to 20.6\% for LOGIS based on \verb|OP+YZsel| compared to \verb|YZsel|, and from 26.3\% to 41.6\% for random forests based on \verb|OP+YZsel| compared to \verb|YZsel| (Table A.6). The variabilities for \verb|OP+All|, \verb|OP+Ysel|, and \verb|OP+YZsel| were close to one another, and one was not uniformly smaller than the other two. In general, regardless of whether OP was incorporated, CART had the largest variability and RMSE among the methods for all scenarios considered. 

In the presence of censoring, OAL and LOGIS based on \verb|OP+YZsel| had the smallest RMSE among the methods under comparison (Table A.18). The inclusion of OP reduced the variability of the effect estimates for all method considered. The bias for LOGIS, CART, and pruned CART decreased when the OP was taken into account, with \verb|OP+YZsel| resulting in the smallest bias for each method. However, we still observed residual bias, especially for the CART and pruned-CART.

\subsubsection{Coverage of 95\% CI}
For the tree-based methods and unpenalized LOGIS, standard errors obtained using the modified bootstrap were close to the corresponding Monte Carlo standard deviations in the scenarios with `sparse' models (regardless of whether the treatment models were linear and/or additive). The modified bootstrap tended to slightly overestimate the variability for LOGIS with LASSO penalty. For example, the ratio of standard errors to Monte Carlo standard deviations ranged from 1.12 to 1.21 for \verb|OP+All| in the scenario with linear associations and `sparse' representation of the models for sample size of 500. The tree-based methods achieved close to nominal coverage of 95\% for \verb|OP+All|, \verb|OP+Ysel|, and \verb|OP+YZsel| for most of the scenarios considered. In the case where the coverage fell below the nominal level, for example the `dense' scenario (Table A.8), the under-coverage was mainly caused by the empirical bias rather than the underestimation of the standard errors.

\section{Data Analysis}
\label{s:Chap3_data}
We applied the algorithms considered in Figure \ref{fig:Chap3_flowchart} and the propensity score estimation methods listed in Table~\ref{tab:Chap3_methods_choices} to a dataset of patients with metastatic castration-resistant prostate cancer (mCRPC) from the Optum Clinformatics Data Mart \cite{ross2021veridical}. Patients who used at least one of the six drugs (docetaxel, abiraterone, enzalutamide, sipuleucel-T, cabazitaxel, and radium-223) approved to treat mCRPC from January 1, 2014, to December 31, 2019 were included in the analytic cohort. We excluded the patients who received cabazitaxel ($n=56$) or radium-223 ($n=28$) as their first-line therapy from our analysis, since there were much fewer samples in these two groups than the other four. The adverse effects of the four drugs for mCRPC were compared, with the outcome being the occurrence of at least one emergency room (ER) visit and hospitalization, respectively, within 180 or 360 days of treatment initiation. In the previous analysis, the working model for the outcome was specified as
\begin{align*}
    \log\frac{P(Y_i=1|\boldsymbol{A}_i,\boldsymbol{B}_i)}{P(Y_i=0|\boldsymbol{A}_i,\boldsymbol{B}_i)} = \boldsymbol{A}_i^T\boldsymbol{\beta}_A+\boldsymbol{B}_i^T\boldsymbol{\beta}_B,
\end{align*}
where $\boldsymbol{A}_i$ contained the sociodemographic factors and other relevant covariates, and $\boldsymbol{B}_i$ contained five pre-existing comorbid conditions, including diabetes, hypertension, arrhythmia, congestive heart failure, and osteoporosis. Specifically, $\boldsymbol{A}_i$ included age, race, education level, household income, geographic region, insurance product type, whether the insurance plan is administrative services only (ASO), metastatic status of cancer, year of first prescription, and provider type \citep{caram2019}. In this analysis, we increased the granularity of the comorbid conditions and considered a list of phenotype codes (phecodes), which are aggregations of the International Classification of Diseases (ICD) codes that represent clinically meaningful phenotypes \citep{denny2010}. We matched the ICD codes in the claims data to the list of phecodes and identified 1042 phecodes as the original reservoir of predictors. These 1042 phecodes represent 16 broad categories of diseases (circulatory system, congenital anomalies, dermatological diseases, endocrine/metabolic diseases, genitourinary diseases, hematopoietic diseases, infectious diseases, injuries and poisonings, mental disorders, musculoskeletal diseases, neoplasms, neurological diseases, respiratory diseases, sense organs, and symptoms). The working model for the outcome became
\begin{align*}
    \log\frac{P(Y_i=1|\boldsymbol{A}_i,\boldsymbol{M}_i)}{P(Y_i=0|\boldsymbol{A}_i,\boldsymbol{M}_i)} = \boldsymbol{A}_i^T\boldsymbol{\beta}_A+\boldsymbol{M}_i^T\boldsymbol{\beta}_M,
\end{align*}
where $\boldsymbol{M}_i$ denotes the list of phecodes of dimension 1042. In the pre-selection step, a LASSO was fitted to select the phecodes that are predictive of the outcome, with the coefficients $\boldsymbol{\beta}_A$ unpenalized. Covariates predictive of the treatment were selected in a similar manner using a multinomial logistic regression with LASSO-type penalty. In other words, covariates in $\boldsymbol{A}_i$ (such as age, race, and household income) are always adjusted for in the treatment and outcome models. The six routes in Figure~\ref{fig:Chap3_flowchart} were followed to obtain six different sets of estimated propensity scores. Note that standard multinomial logistic regression models adjusting for \verb|Ysel| and \verb|YZsel| yielded substantial standard errors for the estimates of ATE (results not shown). Instead, we fitted the models using the LASSO-type penalty (with $\boldsymbol{\beta}_A$ not being penalized) to reduce the variability of the estimates. For censoring, we fitted a Cox model adjusting for $\boldsymbol{A}_i$ and $\boldsymbol{B}_i$, which was a low-dimensional set of covariates. All covariates were coded binary, and the covariates with more than two levels were represented by dummy variables. The standard errors and CIs were constructed using the modified bootstrap procedure. 

\subsection{Data Analysis Results}
The sample sizes for the ER visit cohort and the hospitalization cohort were 7678 and 7709, respectively. The descriptive statistics of the data and the proportions of patients being censored for each treatment group are summarized elsewhere \citep{yu2022inverse}. Specifically, the overall proportions of patients being censored within 180 days and 360 days were 20.8\% and 32.6\%, respectively, for ER visits, and 24.6\% and 40.6\%, respectively, for hospitalization. Sipuleucel-T group had larger percentage of censored patients than the other three groups.

Propensity scores estimated by different routes in Figure~\ref{fig:Chap3_flowchart} were highly correlated for each treatment level (results not shown). In general, we observed larger correlations among propensity scores estimated by LOGIS than those estimated by the tree-based methods. For example, correlations between LOGIS based on \verb|YZsel| and \verb|OP+YZsel| ranged from 0.97 to 1, while the correlations between random forests based on \verb|YZsel| and \verb|OP+YZsel| ranged from 0.93 to 0.99.

The number of phecodes in each disease group selected by the outcome and/or the treatment model in the pre-selection step for each of the two endpoints are reported in Tables A.19-A.22 in the Supplementary Materials. For example, 54 phecodes were selected by the outcome model and 69 were selected by the treatment model, with 12 lying in the intersection for ER visits within 180 days. Among the 12 phecodes predictive of both treatment and outcome, 4 were associated with neoplasm (cancer of prostate, secondary malignancy of respiratory organs, secondary malignant neoplasm, secondary malignant neoplasm of liver), 2 were associated with circulatory system (congestive heart failure NOS and congestive heart failure nonhypertensive), 2 were associated with mental disorders (delirium dementia and amnestic and other cognitive disorders, tobacco use disorder), 1 was associated with endocrine/metabolic system (type 2 diabetes), 1 was associated with hematopoietic system (lymphadenitis), 1 was associated with respiratory system (abnormal findings examination of lungs), and 1 was associated with symptoms (nausea and vomiting). For hospitalization within 180 days, 63 and 79 phecodes were selected by the outcome and treatment model, respectively, and the intersection contained 10 phecodes. Among the 10 phecodes, 7 were overlapped with those identified for ER visits within 180 days (congestive heart failure NOS, congestive heart failure nonhypertensive, lymphadenitis, cancer of prostate, secondary malignant neoplasm, secondary malignant neoplasm of liver, and nausea and vomiting), with 3 additional phecodes associated with neoplasm (cancer of stomach, malignant neoplasm of head, face, and neck, and secondary malignancy of respiratory organs). We also note that a number of phecodes for genitourinary system were identified to be related either to the treatment (e.g., acute cystitis, chronic renal failure (CKD), nephritis, nephrosis, renal sclerosis, other disorders of the kidney and ureters, renal failure, retention of urine, and urinary tract infection) or to the outcome (e.g., chronic kidney disease stage IV, functional disorders of bladder, hyperplasia of prostate, lump or mass in breast, other disorders of prostate, and prostatitis), while the intersection was empty, possibly due to the fine granularity of the phecodes.

\begin{figure}[p]
  \centering
  \includegraphics[width=0.9\textwidth]{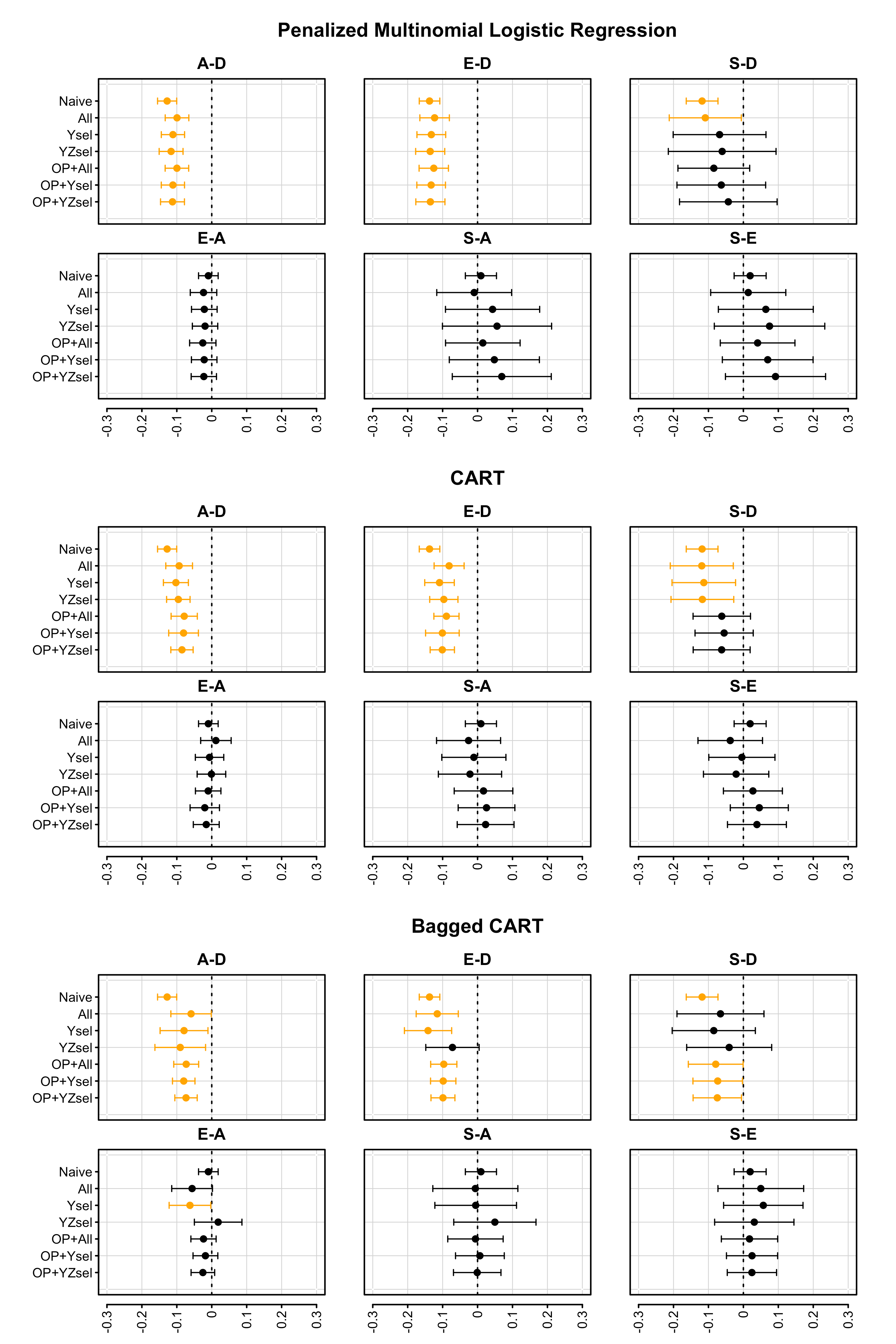}
  \caption[Average treatment effects for ER visits within 180 days of treatment initiation for LOGIS, CART, and bagged CART.]{Average treatment effects for ER visits within 180 days of treatment initiation for LOGIS, CART, and bagged CART. Data were obtained from Optum Clinformative Data Mart. Total sample size was $N=7678$ ($N_A=2757$, $N_D=2311$, $N_E=2043$, $N_S=567$). Confidence intervals that exclude zero are highlighted in orange. Abbreviations: A, abiraterone; D, docetaxel; E, enzalutamide; S, sipuleucel-T.}
\label{fig:Chap3_ER_d180_merged1}
\end{figure}

Figure \ref{fig:Chap3_ER_d180_merged1} showed results for (penalized) LOGIS, CART, and Bagged CART for the 180-day risks differences in ER visits among the four treatment groups. Docetaxel users exhibited significantly higher 180-day risks of at least one ER visit than the users of the two oral drugs (abiraterone and enzalutamide), a finding that is consistent with previous studies \citep{yu2021,tonyali2017}. For example, compared to docetaxel users, 
the 180-day risk of ER visits for abiraterone users is 11.2\% lower (estimated ATE: -0.112 (-0.147,-0.078 )) and for enzalutamide users is 13.5\% lower (estimated ATE: -0.135 (-0.177, -0.093)), from the penalized LOGIS method using \verb|OP+YZsel|. Findings from CART and Bagged CART are consistent with the penalized LOGIS method. The 180-day risk differences between abiraterone and enzalutamide users were generally not significant, except that some of the results yielded by random forests indicated significantly lower risk for the enzalutamide group (Figures \ref{fig:Chap3_ER_d180_merged1} and B.4). For the 360-day time window, enzalutamide users showed significantly lower risk of ER visits in most cases (Figures B.5 and B.6). 

Similar directions of the comparisons among the four treatment groups were observed for 180-day and 360-day risks in hospitalization. In particular, patients who received enzalutamide as their first-line therapy had significantly lower risk of hospitalization than those who received abiraterone, which is consistent with the findings of a study based on a French insurance system database \citep{scailteux2021}.

As was observed in the simulation studies, bagged CART and random forests using \verb|YZsel| as input yielded estimates with large standard errors. In general, incorporating the OP into the treatment model reduced the variability of the estimates and led to narrower 95\% CIs. Greater efficiency gains were noted for the CART family methods compared to LOGIS. For example, for the 360-day risk of ER visits, the ratios of CI widths of \verb|OP+All| over \verb|All| ranged from 0.98 to 1 and from 0.61 to 0.66 for LOGIS and bagged CART, respectively. When the OP were not included in the treatment model, the estimates yielded by LOGIS tended to have lower variance than those produced by CART family methods. For example, for the 360-day risk of ER visits, the ratios of CI widths of bagged CART over CI widths of LOGIS using \verb|All| as input ranged from 1.06 to 2.01. Consistent with what was observed in the simulation studies, the point estimates and confidence widths of \verb|OP+Ysel| and \verb|OP+YZsel| were very close across the propensity score estimation methods.

\section{Discussion}

In this paper, we proposed to incorporate the outcome-covaraite relationship by including predicted outcome probability (OP) in the propensity score model. We examined the traditional multinomial logistic regression method and a set of machine learning techniques combined with OP for estimating ATE. Although the idea of using outcome models to improve efficiency has been explored for regularization methods, how to incorporate outcome-covariate relationship into tree-based machine learning methods has been less studied. Simulation studies show that simultaneous variable selection and propensity score estimation (i.e., methods based on \verb|All|) in a high-dimensional setting led to substantial bias for the LOGIS method, possibly because of the slow convergence rate resulting from the large number of noise variables. Similar pattern was observed for the tree-based methods. We showed that the inclusion of OP can improve the robustness and statistical efficiency of the treatment effect estimators. If the variable selection step had satisfactory performance in terms of identifying the set of important confounders and controlling for the bias, then the benefits of including OP in terms of bias reduction may be minimal. On the other hand, if methods based on \verb|All|, \verb|Ysel|, and \verb|YZsel| produced biased estimates of ATE, then further adjusting for OP in the treatment model can help reduce the bias. OP alleviates the bias by adding back the information about the confounders that are potentially missed by the variable selection procedure.

The LOGIS method (both standard and penalized depending on the dimension of the covariates) outperformed the tree-based machine learning methods in the majority of scenarios, especially when OP is used. The bias for the LOGIS method could be smaller than the tree-based methods even under conditions of nonlinearity and/or nonadditivity. On the other hand, the nonparametric machine learning methods such as bagged CART and random forests could produce less biased estimates than multinomial logistic regression when only the treatment-covariate relationship was considered. The performance of tree-based methods may be improved by optimizing the tuning parameters, such as minimum size of terminal nodes, maximum number of terminal nodes, and number of trees to grow. In addition, standard cross-validation procedure, which focuses on out-of-sample performance, is often used to optimize tuning parameters for accurate predictive performance in practice and may not have desired characteristics for selecting tuning parameters for causal inference (i.e., unbiased treatment effect estimates) \citep{haggstrom2014}. The selection criteria for the tuning parameters in the context of treatment effect estimation could be a topic of future work. 

\section*{Acknowledgment}

This research is supported by NIH (1UG3CA267907), NSF (DMS 1712933) and NIH (R01HG008773).

\section*{Conflict of interest}
Authors state no conflict of interest.

\section*{Data availability statement}

The data that support the findings of this study are available from OptumInsight but restrictions apply to the availability of these data, which were used under license for the current study, and so are not publicly available. Data are however available from the authors upon reasonable request and with permission of OptumInsight (https://ihpi.umich.edu/member-resources/data-and-methods/available-datasets).

\bibliography{references}

\begin{figure}[p]
    \centering
    \includegraphics[width=\textwidth]{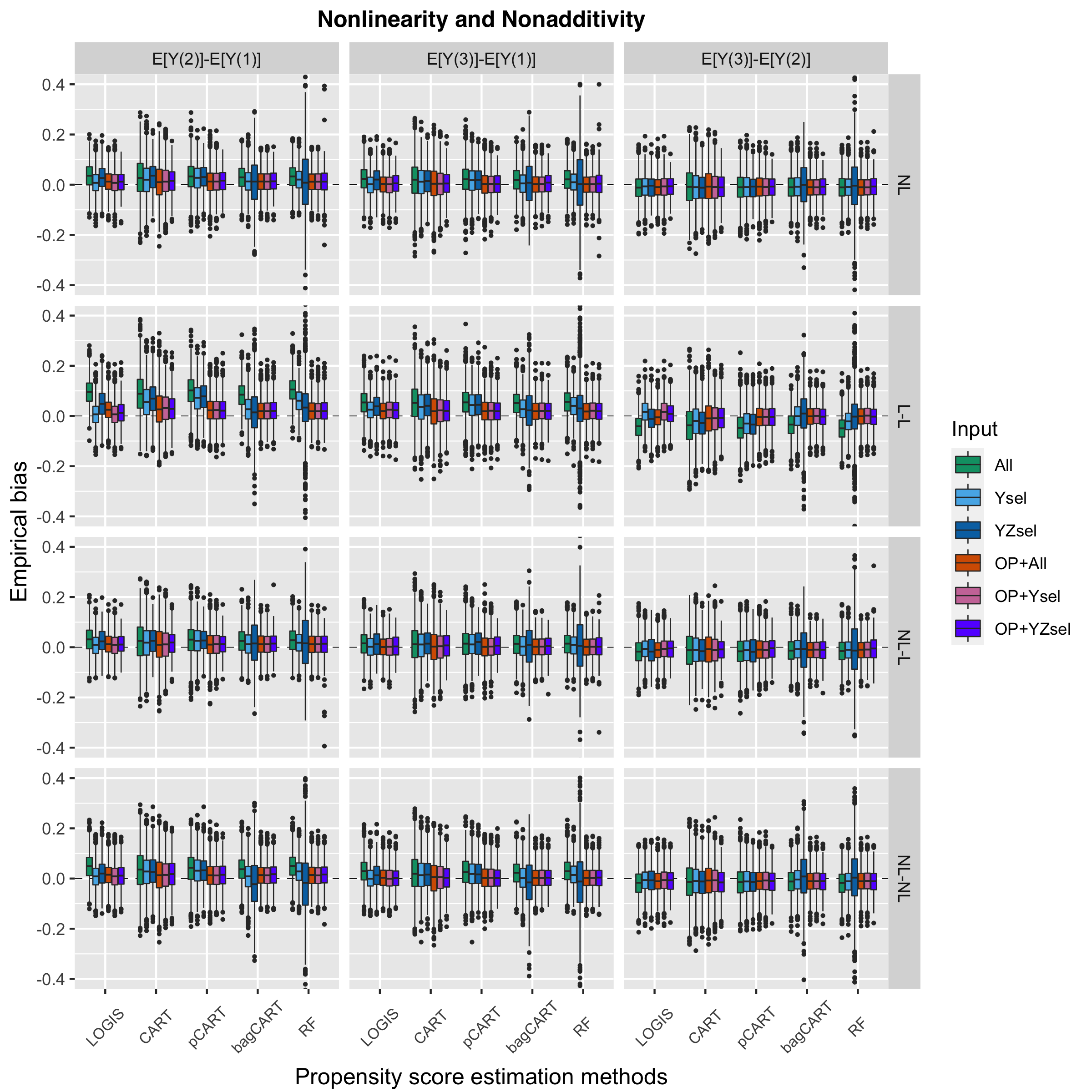}
    \caption[Box plots of empirical bias for 2000 inverse probability weighted estimates for the ATE under scenarios with various degrees of nonlinearity and nonadditivity in the treatment generating model.]{Box plots of empirical bias for 2000 inverse probability weighted estimates for the ATE under scenarios with various degrees of nonlinearity and nonadditivity in the treatment generating model. The rows represent scenarios and columns represent treatment pairs. Each simulated dataset contained 500 samples.}
    \label{fig:Chap3_bias_nonlinear_OP}
\end{figure}

\begin{figure}[p]
  \centering
  \includegraphics[width=0.9\textwidth]{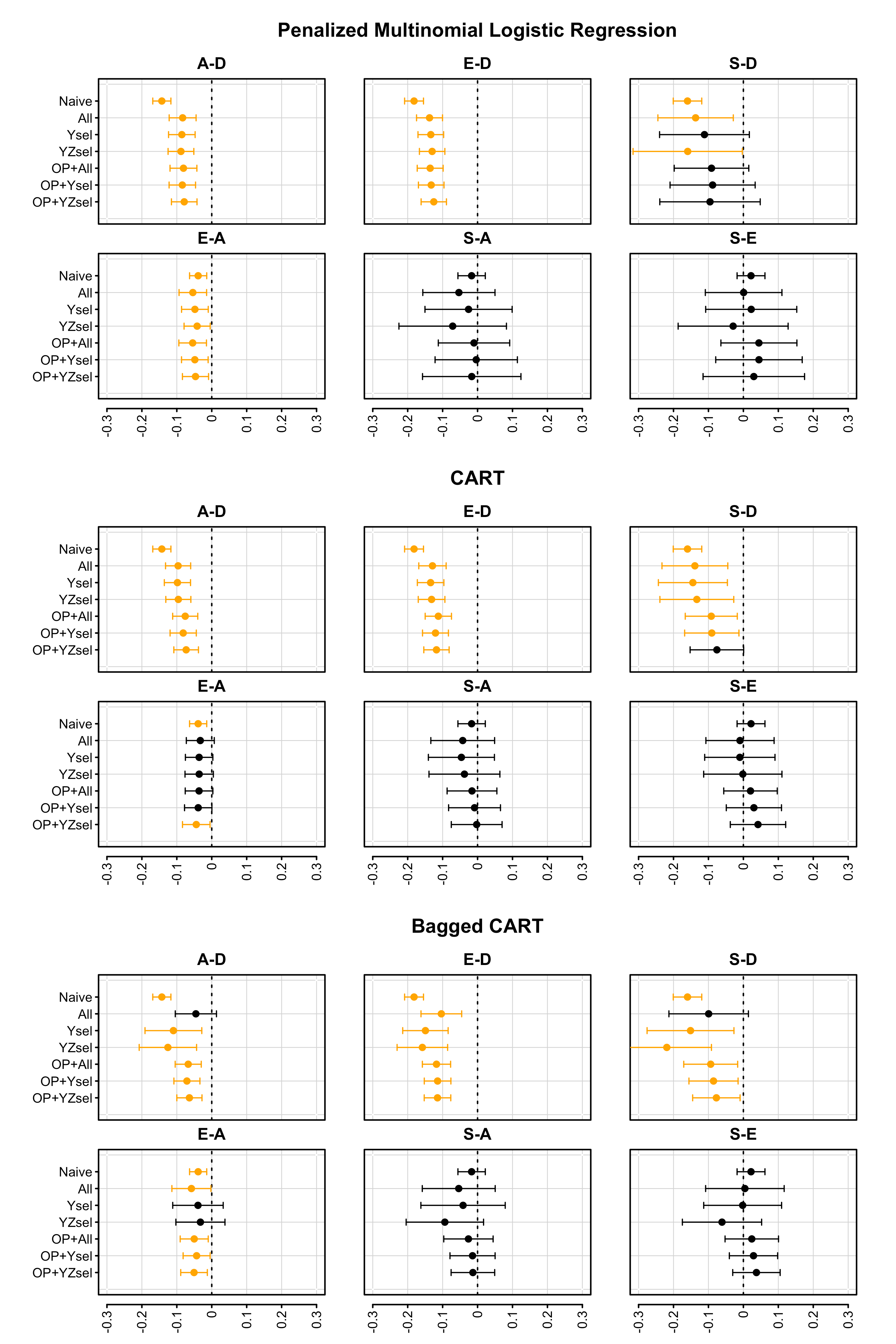}
  \caption[Average treatment effects for hospitalization within 180 days of treatment initiation for LOGIS, CART, and bagged CART.]{Average treatment effects for hospitalization within 180 days of treatment initiation for LOGIS, CART, and bagged CART. Data were obtained from Optum Clinformative Data Mart. Total sample size was $N=7709$ ($N_A=2766$, $N_D=2320$, $N_E=2051$, $N_S=572$. Confidence intervals that exclude zero are highlighted in orange. Abbreviations: A, abiraterone; D, docetaxel; E, enzalutamide; S, sipuleucel-T.}
\label{fig:Chap3_ER_hosp_merged1}
\end{figure}


\section*{Supplementary Materials}

\beginsupplement

\subsection*{Supplementary Tables}

\begin{table}[ht]
\resizebox{\textwidth}{!}{%
\begin{tabular}{@{}lc@{}}
\toprule
\textbf{Scenario} & \textbf{Predictors for Treatment Generating Model} \\ \midrule
\rowcolor[HTML]{EFEFEF} 
Linear main effects only (L)$^\ast$ & $\begin{aligned}[t]&(X_{\mathcal{C}1}, X_{\mathcal{C}2}, X_{\mathcal{C}3},  X_{\mathcal{C}4}, X_{\mathcal{C}5}, X_{\mathcal{Z}1}, X_{\mathcal{Z}2}, X_{\mathcal{Z}3},  X_{\mathcal{Z}4}, X_{\mathcal{Z}5})\end{aligned}$ \\
Nonlinear main effects only (NL) & \begin{tabular}[c]{@{}c@{}}$\begin{aligned}[t](X_{\mathcal{C}1}(X_{\mathcal{C}1}>0), \exp\{X_{\mathcal{C}2}\}^{-1/2}, |X_{\mathcal{C}3}|,  \log|X_{\mathcal{C}4}+1|, |X_{\mathcal{C}5}|^{-1/2},\\ X_{\mathcal{Z}1}, X_{\mathcal{Z}2}, X_{\mathcal{Z}3},  X_{\mathcal{Z}4}^2, \log|X_{\mathcal{Z}5}|)\end{aligned}$\end{tabular} \\
\rowcolor[HTML]{EFEFEF} 
\begin{tabular}[c]{@{}l@{}}Linear main effects \\ and linear interactions (L-L)\end{tabular} & \begin{tabular}[c]{@{}c@{}}$\begin{aligned}[t](X_{\mathcal{C}1}, X_{\mathcal{C}2}, X_{\mathcal{C}3},  X_{\mathcal{C}4}, X_{\mathcal{C}5},\\ X_{\mathcal{Z}1}, X_{\mathcal{Z}2}, X_{\mathcal{Z}3},  X_{\mathcal{Z}4}, X_{\mathcal{Z}5},\\ X_{\mathcal{Z}1}\times X_{\mathcal{Z}2}, X_{\mathcal{Z}1}\times X_{\mathcal{Z}3}, X_{\mathcal{Z}1}\times X_{\mathcal{Z}4}, X_{\mathcal{Z}1}\times X_{\mathcal{Z}5},\\ X_{\mathcal{Z}2}\times X_{\mathcal{Z}3}, X_{\mathcal{Z}2}\times X_{\mathcal{Z}4}, X_{\mathcal{Z}2}\times X_{\mathcal{Z}5}, X_{\mathcal{Z}3}\times X_{\mathcal{Z}4},\\ X_{\mathcal{Z}3}\times X_{\mathcal{Z}5}, X_{\mathcal{Z}4}\times X_{\mathcal{Z}5}, X_{\mathcal{C}3}\times X_{\mathcal{Z}1})\end{aligned}$\end{tabular} \\
\begin{tabular}[c]{@{}l@{}}Non-linear main effects \\ and linear interactions (NL-L)\end{tabular} & \begin{tabular}[c]{@{}c@{}}$\begin{aligned}[t](X_{\mathcal{C}1}(X_{\mathcal{C}1}>0), \exp\{X_{\mathcal{C}2}\}^{-1/2}, |X_{\mathcal{C}3}|,  \log|X_{\mathcal{C}4}+1|, |X_{\mathcal{C}5}|^{-1/2},\\ X_{\mathcal{Z}1}, X_{\mathcal{Z}2}, X_{\mathcal{Z}3},  X_{\mathcal{Z}4}^2, \log|X_{\mathcal{Z}5}|,\\ X_{\mathcal{Z}1}\times X_{\mathcal{Z}2}, X_{\mathcal{Z}1}\times X_{\mathcal{Z}3}, X_{\mathcal{Z}1}\times X_{\mathcal{Z}4}, X_{\mathcal{Z}1}\times X_{\mathcal{Z}5},\\ X_{\mathcal{Z}2}\times X_{\mathcal{Z}3},  X_{\mathcal{Z}2}\times X_{\mathcal{Z}4}, X_{\mathcal{Z}2}\times X_{\mathcal{Z}5}, X_{\mathcal{Z}3}\times X_{\mathcal{Z}4},\\ X_{\mathcal{Z}3}\times X_{\mathcal{Z}5}, X_{\mathcal{Z}4}\times X_{\mathcal{Z}5}, X_{\mathcal{C}3}\times X_{\mathcal{Z}1})\end{aligned}$\end{tabular} \\
\rowcolor[HTML]{EFEFEF} 
\begin{tabular}[c]{@{}l@{}}Non-linear main effects \\ and non-linear interactions (NL-NL)\end{tabular} & \begin{tabular}[c]{@{}c@{}}$\begin{aligned}[t](X_{\mathcal{C}1}(X_{\mathcal{C}1}>0), \sqrt{\exp\{X_{\mathcal{C}2}\}}, |X_{\mathcal{C}3}|,  \log|X_{\mathcal{C}4}+1|, \sqrt{|X_{\mathcal{C}5}|},\\ X_{\mathcal{Z}1}, X_{\mathcal{Z}2}, X_{\mathcal{Z}3},  X_{\mathcal{Z}4}^2, \log|X_{\mathcal{Z}5}|,\\ X_{\mathcal{C}1}(X_{\mathcal{C}1}>0)\times\sqrt{\exp\{X_{\mathcal{C}2}\}}, (X_{\mathcal{C}1}(X_{\mathcal{C}1}>0)\times |X_{\mathcal{C}3}|, \\ X_{\mathcal{C}1}(X_{\mathcal{C}1}>0)\times\log|X_{\mathcal{C}4}+1|,  X_{\mathcal{C}1}(X_{\mathcal{C}1}>0)\times \sqrt{|X_{\mathcal{C}5}|},\\ \sqrt{\exp\{X_{\mathcal{C}2}\}}\times |X_{\mathcal{C}3}|,  \sqrt{\exp\{X_{\mathcal{C}2}\}}\times \log|X_{\mathcal{C}4}+1|, \\ \sqrt{\exp\{X_{\mathcal{C}2}\}}\times\sqrt{|X_{\mathcal{C}5}|}, |X_{\mathcal{C}3}|\times \log|X_{\mathcal{C}4}+1|, \\ |X_{\mathcal{C}3}|\times \sqrt{|X_{\mathcal{C}5}|}, \log|X_{\mathcal{C}4}+1|\times \sqrt{|X_{\mathcal{C}5}|}\\ |X_{\mathcal{C}3}|\times X_{\mathcal{Z}1}\end{aligned}$\end{tabular} \\ \midrule
$^\ast$ The baseline scenario. & \multicolumn{1}{l}{} \\ \bottomrule
\end{tabular}%
}
\caption{Design matrix for the treatment generating models of various degrees of nonlinearity and/or nonadditivity.}
\label{tab:Chap3_Suppl_nonlinear_scenario}
\end{table}

\begin{table}[ht]
\resizebox{\textwidth}{!}{%
\begin{tabular}{rcccccccccccccccccc}
\toprule
\multicolumn{1}{l}{}                     & \multicolumn{3}{c}{\begin{tabular}[c]{@{}c@{}}Bias\\      $\times$1000\end{tabular}} & \multicolumn{3}{c}{\begin{tabular}[c]{@{}c@{}}Empirical SD\\      $\times$1000\end{tabular}} & \multicolumn{3}{c}{\begin{tabular}[c]{@{}c@{}}SE (usual)\\      $\times$1000\end{tabular}} & \multicolumn{3}{c}{\begin{tabular}[c]{@{}c@{}}Coverage (\%)\\  (usual)    \end{tabular}} & \multicolumn{3}{c}{\begin{tabular}[c]{@{}c@{}}SE (modified)\\      $\times$1000\end{tabular}} & \multicolumn{3}{c}{\begin{tabular}[c]{@{}c@{}}Coverage (\%)\\ (modified) \end{tabular}} \\ \cmidrule(lr){2-4} \cmidrule(lr){5-7} \cmidrule(lr){8-10} \cmidrule(lr){11-13} \cmidrule(lr){14-16} \cmidrule(lr){17-19} 
\multicolumn{1}{l}{Estimators}            & 1 vs 2                   & 1 vs 3                  & 2 vs 3                  & 1 vs 2                     & 1 vs 3                     & 2 vs 3                     & 1 vs 2                     & 1 vs 3                    & 2 vs 3                    & 1 vs 2                      & 1 vs 3                     & 2 vs 3                     & 1 vs 2                      & 1 vs 3                     & 2 vs 3                     & 1 vs 2                      & 1 vs 3                     & 2 vs 3  
\\ \toprule
\multicolumn{1}{l}{\textbf{OAL}}         & 95                         & 70                         & -25                        & 37                         & 34                         & 36                         & 39                         & 35                         & 37                         & 96.0                        & 94.9                       & 94.5                       & 38                          & 34                         & 36                         & 94.9                        & 94.9                       & 94.9                       \\
\multicolumn{1}{l}{\textbf{LOGIS}}       &                            &                            &                            &                            &                            &                            &                            &                            &                            &                             &                            &                            &                             &                            &                            &                             &                            &                            \\
All                                      & 72                         & 56                         & -16                        & 37                         & 37                         & 35                         & 40                         & 37                         & 37                         & 55.5                        & 65.9                       & 92.7                       & 41                          & 40                         & 38                         & 59.8                        & 74.2                       & 94.6                       \\
Ysel                                     & -4                         & -2                         & 2                          & 38                         & 35                         & 36                         & 48                         & 41                         & 44                         & 98.2                        & 97.3                       & 97.3                       & 39                          & 35                         & 37                         & 95.4                        & 95.0                       & 94.9                       \\
YZsel                                    & -3                         & -1                         & 2                          & 39                         & 37                         & 38                         & 46                         & 40                         & 43                         & 97.4                        & 96.0                       & 95.9                       & 40                          & 36                         & 38                         & 95.0                        & 95.3                       & 95.4                       \\
OP+All                                   & 14                         & 13                         & -2                         & 37                         & 35                         & 35                         & 40                         & 36                         & 38                         & 95.3                        & 94.1                       & 95.8                       & 43                          & 40                         & 41                         & 96.6                        & 97.6                       & 97.2                       \\
OP+Ysel                                  & -4                         & -2                         & 2                          & 38                         & 35                         & 36                         & 48                         & 41                         & 44                         & 98.2                        & 97.3                       & 97.3                       & 39                          & 35                         & 37                         & 95.4                        & 95.0                       & 94.9                       \\
OP+YZsel                                 & -3                         & -1                         & 2                          & 37                         & 35                         & 36                         & 45                         & 39                         & 42                         & 97.7                        & 96.8                       & 97.0                       & 38                          & 34                         & 37                         & 94.8                        & 95.0                       & 94.8                       \\
\multicolumn{1}{l}{\textbf{CART}}        &                            &                            &                            &                            &                            &                            &                            &                            &                            &                             &                            &                            &                             &                            &                            &                             &                            &                            \\
All                                      & 112                        & 82                         & -30                        & 58                         & 58                         & 55                         & 69                         & 69                         & 65                         & 65.8                        & 81.6                       & 95.7                       & 57                          & 58                         & 55                         & 51.0                        & 70.8                       & 91.9                       \\
Ysel                                     & 61                         & 43                         & -18                        & 50                         & 50                         & 47                         & 63                         & 62                         & 59                         & 87.6                        & 93.5                       & 98.4                       & 52                          & 51                         & 50                         & 78.1                        & 86.8                       & 94.9                       \\
YZsel                                    & 51                         & 38                         & -13                        & 48                         & 47                         & 46                         & 60                         & 59                         & 56                         & 91.9                        & 95.3                       & 97.4                       & 50                          & 49                         & 48                         & 84.3                        & 89.4                       & 95.1                       \\
OP+All                                   & 7                          & 4                          & -3                         & 55                         & 52                         & 56                         & 70                         & 65                         & 71                         & 98.2                        & 98.5                       & 98.3                       & 56                          & 52                         & 56                         & 94.0                        & 94.4                       & 94.4                       \\
OP+Ysel                                  & 13                         & 7                          & -6                         & 45                         & 44                         & 45                         & 61                         & 57                         & 61                         & 98.8                        & 98.6                       & 99.2                       & 46                          & 44                         & 46                         & 94.0                        & 95.1                       & 95.3                       \\
OP+YZsel                                 & 17                         & 11                         & -6                         & 43                         & 41                         & 43                         & 57                         & 54                         & 58                         & 98.4                        & 98.4                       & 99.4                       & 44                          & 41                         & 44                         & 93.5                        & 95.0                       & 94.7                       \\
\multicolumn{1}{l}{\textbf{Pruned CART}} &                            &                            &                            &                            &                            &                            &                            &                            &                            &                             &                            &                            &                             &                            &                            &                             &                            &                            \\
All                                      & 125                        & 94                         & -31                        & 45                         & 43                         & 38                         & 68                         & 68                         & 63                         & 53.0                        & 80.4                       & 99.1                       & 42                          & 42                         & 39                         & 19.4                        & 40.2                       & 88.6                       \\
Ysel                                     & 86                         & 64                         & -22                        & 44                         & 43                         & 38                         & 62                         & 61                         & 57                         & 78.6                        & 88.6                       & 98.5                       & 42                          & 42                         & 39                         & 48.4                        & 66.0                       & 92.8                       \\
YZsel                                    & 75                         & 56                         & -19                        & 46                         & 45                         & 39                         & 60                         & 58                         & 55                         & 80.6                        & 90.7                       & 97.8                       & 43                          & 42                         & 40                         & 60.0                        & 73.4                       & 93.9                       \\
OP+All                                   & -4                         & -1                         & 3                          & 39                         & 36                         & 38                         & 68                         & 62                         & 68                         & 99.9                        & 100.0                      & 99.9                       & 40                          & 36                         & 39                         & 94.0                        & 95.0                       & 95.9                       \\
OP+Ysel                                  & -1                         & -1                         & 1                          & 36                         & 35                         & 37                         & 58                         & 54                         & 59                         & 99.9                        & 99.3                       & 99.5                       & 38                          & 34                         & 37                         & 94.6                        & 94.9                       & 95.6                       \\
OP+YZsel                                 & 1                          & 0                          & 0                          & 37                         & 35                         & 37                         & 55                         & 52                         & 56                         & 99.1                        & 99.5                       & 99.5                       & 38                          & 35                         & 37                         & 94.8                        & 95.4                       & 95.2                       \\
\multicolumn{1}{l}{\textbf{Bagged CART}} &                            &                            &                            &                            &                            &                            &                            &                            &                            &                             &                            &                            &                             &                            &                            &                             &                            &                            \\
All                                      & 109                        & 82                         & -27                        & 38                         & 38                         & 36                         & 31                         & 31                         & 29                         & 10.1                        & 28.2                       & 79.1                       & 40                          & 40                         & 38                         & 24.4                        & 46.9                       & 91.1                       \\
Ysel                                     & -1                         & 0                          & 1                          & 52                         & 46                         & 48                         & 39                         & 36                         & 34                         & 84.8                        & 87.0                       & 85.0                       & 52                          & 48                         & 49                         & 94.4                        & 95.8                       & 96.0                       \\
YZsel                                    & -51                        & -40                        & 11                         & 66                         & 58                         & 59                         & 46                         & 41                         & 39                         & 69.8                        & 74.8                       & 81.0                       & 63                          & 57                         & 60                         & 85.8                        & 86.9                       & 95.2                       \\
OP+All                                   & -4                         & 1                          & 4                          & 37                         & 34                         & 36                         & 62                         & 62                         & 68                         & 99.9                        & 99.9                       & 100.0                      & 38                          & 33                         & 37                         & 94.7                        & 95.0                       & 94.7                       \\
OP+Ysel                                  & -7                         & -2                         & 5                          & 34                         & 32                         & 34                         & 57                         & 60                         & 64                         & 99.9                        & 99.9                       & 100.0                      & 36                          & 32                         & 35                         & 94.3                        & 95.6                       & 94.7                       \\
OP+YZsel                                 & -7                         & -2                         & 5                          & 34                         & 32                         & 34                         & 55                         & 57                         & 60                         & 99.3                        & 100.0                      & 99.9                       & 36                          & 32                         & 35                         & 94.5                        & 95.2                       & 94.6                       \\
\multicolumn{2}{l}{\textbf{Random   Forests}}                         &                            &                            &                            &                            &                            &                            &                            &                            &                             &                            &                            &                             &                            &                            &                             &                            &                            \\
All                                      & 126                        & 95                         & -30                        & 34                         & 35                         & 34                         & 28                         & 28                         & 26                         & 2.5                         & 13.2                       & 73.1                       & 37                          & 38                         & 35                         & 7.0                         & 27.8                       & 88.2                       \\
Ysel                                     & 32                         & 25                         & -7                         & 44                         & 39                         & 37                         & 30                         & 29                         & 27                         & 69.3                        & 77.2                       & 84.5                       & 43                          & 41                         & 41                         & 86.2                        & 91.4                       & 96.7                       \\
YZsel                                    & -71                        & -49                        & 22                         & 86                         & 65                         & 63                         & 34                         & 32                         & 29                         & 48.3                        & 57.3                       & 64.3                       & 62                          & 52                         & 58                         & 73.2                        & 78.6                       & 93.5                       \\
OP+All                                   & -4                         & 1                          & 4                          & 38                         & 35                         & 37                         & 43                         & 39                         & 42                         & 96.6                        & 96.9                       & 96.4                       & 39                          & 34                         & 38                         & 95.3                        & 95.2                       & 94.0                       \\
OP+Ysel                                  & -8                         & -2                         & 6                          & 35                         & 32                         & 35                         & 50                         & 47                         & 52                         & 98.6                        & 99.5                       & 98.7                       & 36                          & 32                         & 36                         & 94.7                        & 95.2                       & 94.7                       \\
OP+YZsel                                 & -8                         & -2                         & 6                          & 35                         & 32                         & 35                         & 56                         & 54                         & 59                         & 99.5                        & 99.8                       & 99.8                       & 36                          & 32                         & 36                         & 94.8                        & 95.6                       & 94.7           \\ \toprule           
\end{tabular}
}
\caption[Standard errors (SE) and coverage of 95\% confidence intervals estimated by usual bootstrap and modified bootstrap for sample size of 1000.]{Standard errors (SE) and coverage of 95\% confidence intervals estimated by usual bootstrap and modified bootstrap for sample size of 1000. The scenario with sparse treatment models was considered. Results were obtained based on 1000 simulated datasets. For each dataset, 200 bootstrap samples were generated.}
\label{tab:Chap3_bootstrap_n1000}
\end{table}

\begin{table}[!p]
\resizebox{\textwidth}{!}{%
\begin{tabular}{@{}rccccccccccccccc@{}}
\toprule
\multicolumn{1}{l}{} & \multicolumn{3}{c}{Bias$\times 1000$} & \multicolumn{3}{c}{MCSD$\times 1000$} & \multicolumn{3}{c}{RMSE$\times 1000$} & \multicolumn{3}{c}{SE$\times 1000$} & \multicolumn{3}{c}{Coverage (\%; modified)} \\ \cmidrule(lr){2-4} \cmidrule(lr){5-7} \cmidrule(lr){8-10} \cmidrule(lr){11-13} \cmidrule(lr){14-16}
\multicolumn{1}{l}{Estimators} & \multicolumn{1}{l}{1 vs 2} & 1 vs 3 & 2 vs 3 & 1 vs 2 & 1 vs 3 & 2 vs 3 & 1 vs 2 & 1 vs 3 & 2 vs 3 & 1 vs 2 & 1 vs 3 & 2 vs 3 & 1 vs 2 & 1 vs 3 & 2 vs 3 \\\midrule
\multicolumn{1}{l}{Naive} & 165 & 124 & -41 & 50 & 52 & 49 & 172 & 135 & 64 & 50 & 52 & 48 & 9.7 & 34.1 & 85.5 \\
\multicolumn{1}{l}{OAL} & 5 & 5 & 0 & 56 & 50 & 52 & 57 & 50 & 52 & 55 & 50 & 52 & 94.1 & 95.6 & 94.2 \\
\multicolumn{1}{l}{LOGIS} &  &  &  &  &  &  &  &  &  &  &  &  &  &  &  \\
Confounder & -1 & 1 & 2 & 58 & 53 & 55 & 58 & 53 & 55 & 57 & 52 & 54 & 94.3 & 94.3 & 94.3 \\
Treatment & 2 & 3 & 1 & 73 & 63 & 65 & 73 & 63 & 65 & 69 & 61 & 62 & 94.1 & 94.2 & 93.4 \\
Outcome & -1 & 0 & 2 & 57 & 49 & 53 & 57 & 49 & 53 & 56 & 50 & 53 & 94.5 & 96.1 & 94.1 \\
\bottomrule
\end{tabular}%
}
\caption[Simulation results for the scenario with sparse models and sample size of 500.]{Simulation results for the scenario with sparse models and sample size of 500. Standard errors were estimated based on 200 bootstrap replications. Results were obtained using 2000 simulated datasets.}
\label{tab:Chap3_Suppl_sparse_n500}
\end{table}

\begin{table}[!p]
\resizebox{\textwidth}{!}{%
\begin{tabular}{@{}rccccccccccccccc@{}}
\toprule
\multicolumn{1}{l}{} & \multicolumn{3}{c}{Bias$\times 1000$} & \multicolumn{3}{c}{MCSD$\times 1000$} & \multicolumn{3}{c}{RMSE$\times 1000$} & \multicolumn{3}{c}{SE$\times 1000$} & \multicolumn{3}{c}{Coverage (\%; modified)} \\ \cmidrule(lr){2-4} \cmidrule(lr){5-7} \cmidrule(lr){8-10} \cmidrule(lr){11-13} \cmidrule(lr){14-16}
\multicolumn{1}{c}{Estimators} & \multicolumn{1}{l}{1 vs 2} & 1 vs 3 & 2 vs 3 & 1 vs 2 & 1 vs 3 & 2 vs 3 & 1 vs 2 & 1 vs 3 & 2 vs 3 & 1 vs 2 & 1 vs 3 & 2 vs 3 & 1 vs 2 & 1 vs 3 & 2 vs 3 \\ \midrule
\multicolumn{1}{l}{\textbf{Naive}} & 165 & 125 & -41 & 36 & 37 & 33 & 169 & 130 & 53 & 36 & 37 & 34 & 0.5 & 8.3 & 78.4 \\
\multicolumn{1}{l}{\textbf{OAL}} & -2 & 0 & 2 & 38 & 33 & 36 & 38 & 33 & 36 & 38 & 34 & 36 & 94.9 & 94.9 & 94.9 \\
\multicolumn{1}{l}{\textbf{LOGIS}} &  &  &  &  &  &  &  &  &  &  &  &  &  &  &  \\
Confounder & -3 & 0 & 2 & 40 & 36 & 37 & 40 & 36 & 37 & 40 & 36 & 38 & 94.5 & 95.0 & 95.2 \\
Treatment & -2 & 0 & 1 & 50 & 43 & 43 & 50 & 43 & 43 & 49 & 43 & 43 & 94.2 & 94.7 & 94.4 \\
Outcome & -3 & 0 & 2 & 38 & 33 & 36 & 38 & 33 & 36 & 38 & 34 & 37 & 94.8 & 95.1 & 94.9 \\
\bottomrule
\end{tabular}%
}
\caption[Simulation results for the scenario with sparse models and sample size of 1000.]{Simulation results for the scenario with sparse models and sample size of 1000. Standard errors were estimated based on 200 bootstrap replications. Results were obtained using 2000 simulated datasets.}
\label{tab:Chap3_Suppl_sparse_n1000}
\end{table}

\begin{table}[h]
\resizebox{\textwidth}{!}{%
\begin{tabular}{@{}rccccccccccccccc@{}}
\toprule
\multicolumn{1}{l}{} & \multicolumn{3}{c}{Bias$\times 1000$} & \multicolumn{3}{c}{MCSD$\times 1000$} & \multicolumn{3}{c}{RMSE$\times 1000$} & \multicolumn{3}{c}{SE$\times 1000$} & \multicolumn{3}{c}{Coverage (\%; modified)} \\ \cmidrule(lr){2-4} \cmidrule(lr){5-7} \cmidrule(lr){8-10} \cmidrule(lr){11-13} \cmidrule(lr){14-16}
\multicolumn{1}{c}{Estimators} & \multicolumn{1}{l}{1 vs 2} & 1 vs 3 & 2 vs 3 & 1 vs 2 & 1 vs 3 & 2 vs 3 & 1 vs 2 & 1 vs 3 & 2 vs 3 & 1 vs 2 & 1 vs 3 & 2 vs 3 & 1 vs 2 & 1 vs 3 & 2 vs 3 \\ \midrule
\multicolumn{1}{l}{\textbf{Naive}} & 167 & 125 & -41 & 26 & 27 & 24 & 169 & 128 & 48 & 25 & 26 & 24 & 0 & 0.4 & 59.9 \\
\multicolumn{1}{l}{\textbf{OAL}} & -1 & 0 & 1 & 27 & 24 & 26 & 27 & 24 & 26 & 27 & 24 & 26 & 95.0 & 94.7 & 95.2 \\
\multicolumn{1}{l}{\textbf{LOGIS}} &  &  &  &  &  &  &  &  &  &  &  &  &  &  &  \\
Confounder & 0 & 1 & 1 & 28 & 26 & 27 & 28 & 26 & 27 & 28 & 26 & 27 & 95.1 & 94.6 & 95.2 \\
Treatment & 0 & 0 & 0 & 35 & 31 & 31 & 35 & 31 & 31 & 34 & 30 & 30 & 93.8 & 94.3 & 94.8 \\
Outcome & 0 & 0 & 1 & 27 & 24 & 26 & 27 & 24 & 26 & 27 & 24 & 26 & 94.8 & 94.7 & 95.0 \\
\bottomrule
\end{tabular}%
}
\caption[Simulation results for the scenario with sparse models and sample size of 2000.]{Simulation results for the scenario with sparse models and sample size of 2000. Standard errors were estimated based on 200 bootstrap replications. Results were obtained using 2000 simulated datasets.}
\label{tab:Chap3_Suppl_sparse_n2000}
\end{table}

\begin{table}[h]
\resizebox{\textwidth}{!}{%
\begin{tabular}{@{}rccccccccccccccc@{}}
\toprule
\multicolumn{1}{l}{} & \multicolumn{3}{c}{Bias$\times 1000$} & \multicolumn{3}{c}{MCSD$\times 1000$} & \multicolumn{3}{c}{RMSE$\times 1000$} & \multicolumn{3}{c}{SE$\times 1000$} & \multicolumn{3}{c}{Coverage (\%; modified)} \\ \cmidrule(lr){2-4} \cmidrule(lr){5-7} \cmidrule(lr){8-10} \cmidrule(lr){11-13} \cmidrule(lr){14-16}
\multicolumn{1}{c}{Estimators} & \multicolumn{1}{l}{1 vs 2} & 1 vs 3 & 2 vs 3 & 1 vs 2 & 1 vs 3 & 2 vs 3 & 1 vs 2 & 1 vs 3 & 2 vs 3 & 1 vs 2 & 1 vs 3 & 2 vs 3 & 1 vs 2 & 1 vs 3 & 2 vs 3 \\ \midrule
\multicolumn{1}{l}{\textbf{Naive}} & 141 & 108 & -32 & 49 & 48 & 44 & 149 & 119 & 54 & 48 & 50 & 43 & 17.7 & 40.6 & 88.4 \\
\multicolumn{1}{l}{\textbf{OAL}} & 24 & 19 & -5 & 53 & 48 & 48 & 58 & 52 & 48 & 50 & 47 & 47 & 90.2 & 92.6 & 94.1 \\
\multicolumn{1}{l}{\textbf{LOGIS}} &  &  &  &  &  &  &  &  &  &  &  &  &  &  &  \\
Confounder & 0 & 2 & 1 & 54 & 48 & 52 & 54 & 48 & 52 & 53 & 49 & 51 & 94.4 & 95.3 & 95.0 \\
Treatment & 1 & 2 & 2 & 64 & 54 & 59 & 64 & 54 & 59 & 64 & 56 & 58 & 95.0 & 95.3 & 94.3 \\
Outcome & 0 & 2 & 1 & 54 & 46 & 52 & 54 & 46 & 52 & 54 & 48 & 52 & 94.8 & 95.9 & 94.4 \\
\bottomrule
\end{tabular}%
}
\caption[Simulation results for the scenario with moderately sparse models and sample size of 500.]{Simulation results for the scenario with moderately sparse models and sample size of 500. Standard errors were estimated based on 200 bootstrap replications. Results were obtained using 2000 simulated datasets.}
\label{tab:Chap3_Suppl_mod_n500}
\end{table}

\begin{table}[h]
\resizebox{\textwidth}{!}{%
\begin{tabular}{@{}rccccccccccccccc@{}}
\toprule
\multicolumn{1}{l}{} & \multicolumn{3}{c}{Bias$\times 1000$} & \multicolumn{3}{c}{MCSD$\times 1000$} & \multicolumn{3}{c}{RMSE$\times 1000$} & \multicolumn{3}{c}{SE$\times 1000$} & \multicolumn{3}{c}{Coverage (\%; modified)} \\ \midrule
\multicolumn{1}{c}{Estimators} & \multicolumn{1}{l}{1 vs 2} & 1 vs 3 & 2 vs 3 & 1 vs 2 & 1 vs 3 & 2 vs 3 & 1 vs 2 & 1 vs 3 & 2 vs 3 & 1 vs 2 & 1 vs 3 & 2 vs 3 & 1 vs 2 & 1 vs 3 & 2 vs 3 \\ \midrule
\multicolumn{1}{l}{\textbf{Naive}} & 142 & 109 & -33 & 34 & 35 & 30 & 146 & 114 & 45 & 34 & 35 & 31 & 1.5 & 12.2 & 81.0 \\
\multicolumn{1}{l}{\textbf{OAL}} & 3 & 2 & -1 & 35 & 32 & 34 & 35 & 32 & 34 & 36 & 32 & 34 & 95.9 & 95.3 & 95.6 \\
\multicolumn{1}{l}{\textbf{LOGIS}} &  &  &  &  &  &  &  &  &  &  &  &  &  &  &  \\
Confounder & 1 & 0 & -1 & 36 & 33 & 35 & 36 & 33 & 35 & 37 & 33 & 36 & 95.3 & 94.8 & 95.3 \\
Treatment & 1 & 0 & -1 & 43 & 37 & 39 & 43 & 37 & 39 & 43 & 37 & 39 & 94.8 & 94.3 & 95.2 \\
Outcome & 0 & 0 & 0 & 35 & 32 & 34 & 35 & 32 & 34 & 36 & 32 & 35 & 96.4 & 95.4 & 95.8 \\
\bottomrule
\end{tabular}%
}
\caption[Simulation results for the scenario with moderately sparse models and sample size of 1000.]{Simulation results for the scenario with moderately sparse models and sample size of 1000. Standard errors were estimated based on 200 bootstrap replications. Results were obtained using 2000 simulated datasets.}
\label{tab:Chap3_Suppl_mod_n1000}
\end{table}

\begin{table}[h]
\resizebox{\textwidth}{!}{%
\begin{tabular}{@{}rccccccccccccccc@{}}
\toprule
\multicolumn{1}{l}{} & \multicolumn{3}{c}{Bias$\times 1000$} & \multicolumn{3}{c}{MCSD$\times 1000$} & \multicolumn{3}{c}{RMSE$\times 1000$} & \multicolumn{3}{c}{SE$\times 1000$} & \multicolumn{3}{c}{Coverage (\%; modified)} \\ \cmidrule(lr){2-4} \cmidrule(lr){5-7} \cmidrule(lr){8-10} \cmidrule(lr){11-13} \cmidrule(lr){14-16}
\multicolumn{1}{c}{Estimators} & \multicolumn{1}{l}{1 vs 2} & 1 vs 3 & 2 vs 3 & 1 vs 2 & 1 vs 3 & 2 vs 3 & 1 vs 2 & 1 vs 3 & 2 vs 3 & 1 vs 2 & 1 vs 3 & 2 vs 3 & 1 vs 2 & 1 vs 3 & 2 vs 3 \\ \midrule
\multicolumn{1}{l}{\textbf{Naive}} & 113 & 87 & -26 & 44 & 47 & 38 & 121 & 99 & 46 & 44 & 46 & 38 & 27.8 & 52.5 & 90.2 \\
\multicolumn{1}{l}{\textbf{OAL}} & 51 & 37 & -14 & 48 & 47 & 41 & 70 & 60 & 43 & 44 & 43 & 41 & 76.4 & 84.1 & 93.4 \\
\multicolumn{1}{l}{\textbf{LOGIS}} &  &  &  &  &  &  &  &  &  &  &  &  &  &  &  \\
Confounder & 2 & 1 & -1 & 48 & 45 & 47 & 48 & 45 & 47 & 51 & 46 & 50 & 95.3 & 95.0 & 95.0 \\
Treatment & 2 & 1 & -2 & 57 & 51 & 54 & 57 & 51 & 54 & 65 & 57 & 61 & 97.1 & 96.9 & 96.9 \\
Outcome & 1 & 0 & -1 & 48 & 45 & 48 & 48 & 45 & 48 & 56 & 50 & 54 & 97.5 & 96.5 & 97.2 \\
\bottomrule
\end{tabular}%
}
\caption[Simulation results for the scenario with dense models and sample size of 500.]{Simulation results for the scenario with dense models and sample size of 500. Standard errors were estimated based on 200 bootstrap replications. Results were obtained using 2000 simulated datasets.}
\label{tab:Chap3_Suppl_dense_n500}
\end{table}

\begin{table}[h]
\resizebox{\textwidth}{!}{%
\begin{tabular}{@{}rccccccccccccccc@{}}
\toprule
\multicolumn{1}{l}{} & \multicolumn{3}{c}{Bias$\times 1000$} & \multicolumn{3}{c}{MCSD$\times 1000$} & \multicolumn{3}{c}{RMSE$\times 1000$} & \multicolumn{3}{c}{SE$\times 1000$} & \multicolumn{3}{c}{Coverage (\%; modified)} \\ \cmidrule(lr){2-4} \cmidrule(lr){5-7} \cmidrule(lr){8-10} \cmidrule(lr){11-13} \cmidrule(lr){14-16}
\multicolumn{1}{c}{Estimators} & \multicolumn{1}{l}{1 vs 2} & 1 vs 3 & 2 vs 3 & 1 vs 2 & 1 vs 3 & 2 vs 3 & 1 vs 2 & 1 vs 3 & 2 vs 3 & 1 vs 2 & 1 vs 3 & 2 vs 3 & 1 vs 2 & 1 vs 3 & 2 vs 3 \\ \midrule
\multicolumn{1}{l}{\textbf{Naive}} & 111 & 87 & -24 & 32 & 32 & 27 & 116 & 93 & 36 & 31 & 32 & 27 & 5.8 & 21.6 & 86.3 \\
\multicolumn{1}{l}{\textbf{OAL}} & 12 & 10 & -2 & 32 & 30 & 31 & 34 & 31 & 31 & 32 & 30 & 31 & 93.2 & 94.1 & 94.8 \\
\multicolumn{1}{l}{\textbf{LOGIS}} &  &  &  &  &  &  &  &  &  &  &  &  &  &  &  \\
Confounder & -1 & 0 & 1 & 33 & 30 & 33 & 33 & 30 & 33 & 34 & 31 & 33 & 94.5 & 95.1 & 94.5 \\
Treatment & -1 & 0 & 0 & 38 & 33 & 36 & 38 & 33 & 36 & 39 & 34 & 36 & 95.0 & 95.1 & 94.3 \\
Outcome & -1 & 0 & 1 & 32 & 29 & 33 & 32 & 29 & 33 & 34 & 30 & 33 & 95.4 & 95.5 & 94.6 \\
\bottomrule
\end{tabular}%
}
\caption[Simulation results for the scenario with dense models and sample size of 1000.]{Simulation results for the scenario with dense models and sample size of 1000. Standard errors were estimated based on 200 bootstrap replications. Results were obtained using 2000 simulated datasets.}
\label{tab:Chap3_Suppl_dense_n1000}
\end{table}

\begin{table}[h]
\resizebox{\textwidth}{!}{%
\begin{tabular}{@{}rccccccccccccccc@{}}
\toprule
\multicolumn{1}{l}{} & \multicolumn{3}{c}{Bias$\times 1000$} & \multicolumn{3}{c}{MCSD$\times 1000$} & \multicolumn{3}{c}{RMSE$\times 1000$} & \multicolumn{3}{c}{SE$\times 1000$} & \multicolumn{3}{c}{Coverage (\%; modified)} \\
\cmidrule(lr){2-4} \cmidrule(lr){5-7} \cmidrule(lr){8-10} \cmidrule(lr){11-13} \cmidrule(lr){14-16}
\multicolumn{1}{c}{Estimators} & \multicolumn{1}{l}{1 vs 2} & 1 vs 3 & 2 vs 3 & 1 vs 2 & 1 vs 3 & 2 vs 3 & 1 vs 2 & 1 vs 3 & 2 vs 3 & 1 vs 2 & 1 vs 3 & 2 vs 3 & 1 vs 2 & 1 vs 3 & 2 vs 3 \\ \midrule
\multicolumn{1}{l}{\textbf{Naive}} & 38 & 26 & -13 & 53 & 52 & 53 & 65 & 58 & 55 & 54 & 54 & 54 & 90.3 & 92.4 & 94.8 \\
\multicolumn{1}{l}{\textbf{OAL}} & 10 & 3 & -7 & 47 & 46 & 47 & 48 & 47 & 48 & 55 & 55 & 55 & 97.4 & 97.2 & 97.2 \\
\multicolumn{1}{l}{\textbf{LOGIS}} &  &  &  &  &  &  &  &  &  &  &  &  &  &  &  \\
Confounder & 10 & 7 & -3 & 50 & 49 & 49 & 51 & 49 & 49 & 51 & 49 & 50 & 95.2 & 94.7 & 95.7 \\
Treatment & -1 & 0 & 1 & 51 & 49 & 50 & 51 & 49 & 50 & 52 & 50 & 52 & 95.7 & 95.1 & 95.3 \\
Outcome & -1 & 0 & 1 & 48 & 46 & 47 & 48 & 46 & 47 & 49 & 47 & 48 & 95.3 & 95.8 & 96.0 \\
\bottomrule
\end{tabular}%
}
\caption[Simulation results for the scenario with nonlinear main effects and no interactions.]{Simulation results for the scenario with nonlinear main effects and no interactions. Each simulated dataset contained 500 subjects. Standard errors were estimated based on 200 bootstrap replications. Results were obtained using 2000 simulated datasets.}
\label{tab:Chap3_Suppl_NL_n500}
\end{table}

\begin{table}[h]
\resizebox{\textwidth}{!}{%
\begin{tabular}{@{}rccccccccccccccc@{}}
\toprule
\multicolumn{1}{l}{} & \multicolumn{3}{c}{Bias$\times 1000$} & \multicolumn{3}{c}{MCSD$\times 1000$} & \multicolumn{3}{c}{RMSE$\times 1000$} & \multicolumn{3}{c}{SE$\times 1000$} & \multicolumn{3}{c}{Coverage (\%; modified)} \\
\multicolumn{1}{c}{Estimators} & \multicolumn{1}{l}{1 vs 2} & 1 vs 3 & 2 vs 3 & 1 vs 2 & 1 vs 3 & 2 vs 3 & 1 vs 2 & 1 vs 3 & 2 vs 3 & 1 vs 2 & 1 vs 3 & 2 vs 3 & 1 vs 2 & 1 vs 3 & 2 vs 3 \\ \midrule
\multicolumn{1}{l}{\textbf{Naive}} & 41 & 27 & -15 & 38 & 37 & 38 & 56 & 45 & 41 & 38 & 38 & 38 & 81.0 & 89.6 & 93.8 \\
\multicolumn{1}{l}{\textbf{OAL}} & 6 & 2 & -4 & 34 & 32 & 33 & 34 & 32 & 33 & 39 & 38 & 39 & 97.4 & 98.2 & 98.0 \\
\multicolumn{1}{l}{\textbf{LOGIS}} &  &  &  &  &  &  &  &  &  &  &  &  &  &  &  \\
Confounder & 13 & 8 & -5 & 36 & 34 & 35 & 38 & 35 & 35 & 36 & 34 & 35 & 93.7 & 93.8 & 94.6 \\
Treatment & 2 & 0 & -2 & 36 & 34 & 35 & 36 & 34 & 35 & 36 & 35 & 35 & 95.0 & 94.8 & 95.0 \\
Outcome & 1 & 0 & -2 & 34 & 32 & 33 & 34 & 32 & 33 & 34 & 32 & 33 & 95.2 & 95.3 & 94.9 \\
\bottomrule
\end{tabular}%
}
\caption[Simulation results for the scenario with nonlinear main effects and no interactions.]{Simulation results for the scenario with nonlinear main effects and no interactions. Each simulated dataset contained 1000 subjects. Standard errors were estimated based on 200 bootstrap replications. Results were obtained using 2000 simulated datasets.}
\label{tab:Chap3_Suppl_NL_n1000}
\end{table}

\begin{table}[h]
\resizebox{\textwidth}{!}{%
\begin{tabular}{@{}rccccccccccccccc@{}}
\toprule
\multicolumn{1}{l}{} & \multicolumn{3}{c}{Bias$\times 1000$} & \multicolumn{3}{c}{MCSD$\times 1000$} & \multicolumn{3}{c}{RMSE$\times 1000$} & \multicolumn{3}{c}{SE$\times 1000$} & \multicolumn{3}{c}{Coverage (\%; modified)} \\
\multicolumn{1}{c}{Estimators} & \multicolumn{1}{l}{1 vs 2} & 1 vs 3 & 2 vs 3 & 1 vs 2 & 1 vs 3 & 2 vs 3 & 1 vs 2 & 1 vs 3 & 2 vs 3 & 1 vs 2 & 1 vs 3 & 2 vs 3 & 1 vs 2 & 1 vs 3 & 2 vs 3 \\ \midrule
\multicolumn{1}{l}{\textbf{Naive}} & 128 & 62 & -66 & 54 & 54 & 53 & 139 & 82 & 84 & 54 & 54 & 54 & 35.1 & 78.6 & 78.0 \\
\multicolumn{1}{l}{\textbf{OAL}} & 33 & 31 & -2 & 51 & 48 & 48 & 61 & 57 & 48 & 56 & 55 & 55 & 92.4 & 93.3 & 96.8 \\
\multicolumn{1}{l}{\textbf{LOGIS}} &  &  &  &  &  &  &  &  &  &  &  &  &  &  &  \\
Confounder & 8 & 31 & 23 & 52 & 50 & 50 & 52 & 59 & 55 & 53 & 51 & 51 & 95.6 & 91.1 & 92.2 \\
Treatment & -6 & 23 & 29 & 56 & 52 & 53 & 56 & 57 & 60 & 58 & 54 & 54 & 95.4 & 94.0 & 91.1 \\
Outcome & -3 & 24 & 27 & 49 & 46 & 47 & 49 & 52 & 55 & 51 & 48 & 49 & 95.6 & 92.8 & 91.3 \\
\bottomrule
\end{tabular}%
}
\caption[Simulation results for the scenario with linear main effects and linear interactions for sample size of 500.]{Simulation results for the scenario with linear main effects and linear interactions. Each simulated dataset contained 500 subjects. Standard errors were estimated based on 200 bootstrap replications. Results were obtained using 2000 simulated datasets.}
\label{tab:Chap3_Suppl_L-L_n500}
\end{table}

\begin{table}[]
\resizebox{\textwidth}{!}{%
\begin{tabular}{@{}rccccccccccccccc@{}}
\toprule
\multicolumn{1}{l}{} & \multicolumn{3}{c}{Bias$\times 1000$} & \multicolumn{3}{c}{MCSD$\times 1000$} & \multicolumn{3}{c}{RMSE$\times 1000$} & \multicolumn{3}{c}{SE$\times 1000$} & \multicolumn{3}{c}{Coverage (\%; modified)} \\
\multicolumn{1}{c}{Estimators} & \multicolumn{1}{l}{1 vs 2} & 1 vs 3 & 2 vs 3 & 1 vs 2 & 1 vs 3 & 2 vs 3 & 1 vs 2 & 1 vs 3 & 2 vs 3 & 1 vs 2 & 1 vs 3 & 2 vs 3 & 1 vs 2 & 1 vs 3 & 2 vs 3 \\ \midrule
\multicolumn{1}{l}{\textbf{Naive}} & 128 & 62 & -66 & 38 & 38 & 38 & 133 & 73 & 76 & 38 & 38 & 38 & 8.7 & 63.2 & 59.2 \\
\multicolumn{1}{l}{\textbf{OAL}} & 23 & 31 & 8 & 39 & 34 & 36 & 46 & 46 & 36 & 39 & 39 & 39 & 90.6 & 90.1 & 96.6 \\
\multicolumn{1}{l}{\textbf{LOGIS}} &  &  &  &  &  &  &  &  &  &  &  &  &  &  &  \\
Confounder & 8 & 33 & 25 & 37 & 36 & 35 & 38 & 48 & 43 & 37 & 36 & 36 & 94.4 & 85.0 & 88.3 \\
Treatment & -7 & 24 & 32 & 39 & 37 & 37 & 40 & 44 & 48 & 40 & 37 & 37 & 94.8 & 89.2 & 85.7 \\
Outcome & -4 & 25 & 29 & 35 & 33 & 33 & 35 & 42 & 44 & 35 & 33 & 34 & 94.7 & 88.1 & 85.6 \\
All & 71 & 47 & -24 & 36 & 36 & 35 & 80 & 60 & 42 & 40 & 39 & 39 & 57.5 & 78.4 & 93.4 \\
Ysel & 0 & 25 & 25 & 35 & 33 & 33 & 35 & 42 & 41 & 39 & 39 & 39 & 96.9 & 92.7 & 93.1 \\
YZsel & 16 & 32 & 16 & 40 & 36 & 37 & 43 & 48 & 40 & 37 & 36 & 36 & 91.6 & 86.9 & 91.0 \\
OP + All & 15 & 22 & 7 & 33 & 33 & 32 & 37 & 40 & 33 & 40 & 39 & 39 & 96.4 & 94.2 & 97.2 \\
OP + Ysel & 0 & 25 & 25 & 35 & 33 & 33 & 35 & 42 & 41 & 33 & 33 & 33 & 94.0 & 87.9 & 87.4 \\
OP + YZsel & 3 & 25 & 22 & 35 & 33 & 33 & 35 & 42 & 40 & 35 & 33 & 33 & 94.5 & 88.2 & 89.0 \\
\multicolumn{1}{l}{\textbf{CART}} &  &  &  &  &  &  &  &  &  &  &  &  &  &  &  \\
Confounder & 46 & 34 & -12 & 52 & 50 & 49 & 69 & 60 & 50 & 53 & 51 & 50 & 86.6 & 91.0 & 94.1 \\
Treatment & 56 & 38 & -18 & 54 & 53 & 49 & 78 & 65 & 52 & 55 & 53 & 52 & 82.5 & 88.2 & 94.3 \\
Outcome & 47 & 34 & -14 & 52 & 50 & 48 & 70 & 61 & 50 & 54 & 53 & 52 & 86.4 & 91.0 & 95.8 \\
All & 77 & 47 & -30 & 59 & 58 & 58 & 97 & 75 & 65 & 60 & 58 & 58 & 74.1 & 87.1 & 92.2 \\
Ysel & 52 & 36 & -16 & 53 & 52 & 50 & 74 & 63 & 52 & 55 & 53 & 52 & 85.1 & 89.7 & 94.8 \\
YZsel & 52 & 36 & -16 & 51 & 48 & 47 & 73 & 60 & 50 & 52 & 51 & 50 & 83.9 & 89.3 & 94.7 \\
OP + All & 21 & 18 & -3 & 52 & 52 & 51 & 57 & 55 & 51 & 53 & 52 & 52 & 93.6 & 93.4 & 94.6 \\
OP + Ysel & 29 & 23 & -6 & 47 & 46 & 45 & 55 & 52 & 46 & 48 & 46 & 46 & 90.6 & 90.8 & 94.7 \\
OP + YZsel & 28 & 22 & -6 & 45 & 43 & 43 & 53 & 48 & 44 & 45 & 43 & 43 & 90.4 & 91.4 & 94.2 \\
\multicolumn{1}{l}{\textbf{Pruned CART}} &  &  &  &  &  &  &  &  &  &  &  &  &  &  &  \\
Confounder & 53 & 35 & -18 & 46 & 43 & 42 & 70 & 55 & 46 & 45 & 44 & 43 & 76.2 & 86.7 & 93.2 \\
Treatment & 70 & 43 & -28 & 49 & 45 & 43 & 86 & 62 & 51 & 46 & 44 & 44 & 63.5 & 82.6 & 89.8 \\
Outcome & 58 & 35 & -22 & 46 & 42 & 41 & 74 & 55 & 47 & 44 & 43 & 43 & 71.6 & 86.6 & 91.8 \\
All & 92 & 50 & -41 & 46 & 43 & 43 & 103 & 66 & 59 & 44 & 42 & 42 & 44.0 & 76.9 & 82.2 \\
Ysel & 63 & 38 & -25 & 46 & 43 & 43 & 78 & 57 & 49 & 44 & 43 & 43 & 68.4 & 84.9 & 90.7 \\
YZsel & 60 & 37 & -22 & 46 & 42 & 41 & 75 & 56 & 47 & 44 & 43 & 42 & 71.0 & 86.1 & 92.2 \\
OP + All & 15 & 19 & 4 & 37 & 36 & 35 & 40 & 41 & 36 & 37 & 36 & 36 & 93.6 & 91.3 & 94.7 \\
OP + Ysel & 18 & 20 & 2 & 38 & 36 & 36 & 42 & 42 & 36 & 37 & 36 & 36 & 92.1 & 90.6 & 94.8 \\
OP + YZsel & 18 & 20 & 2 & 38 & 36 & 36 & 42 & 41 & 36 & 37 & 36 & 35 & 91.7 & 90.8 & 94.0 \\
\multicolumn{1}{l}{\textbf{Bagged CART}} &  &  &  &  &  &  &  &  &  &  &  &  &  &  &  \\
Confounder & -5 & 12 & 16 & 53 & 51 & 48 & 53 & 52 & 50 & 53 & 51 & 49 & 93.9 & 93.8 & 94.0 \\
Treatment & 19 & 28 & 8 & 46 & 44 & 42 & 50 & 52 & 43 & 48 & 46 & 46 & 93.8 & 91.2 & 95.5 \\
Outcome & 9 & 20 & 11 & 43 & 42 & 39 & 44 & 46 & 41 & 47 & 45 & 45 & 95.8 & 94.0 & 96.2 \\
All & 74 & 48 & -26 & 38 & 37 & 36 & 83 & 61 & 45 & 41 & 40 & 40 & 55.4 & 79.1 & 92.2 \\
Ysel & 21 & 25 & 4 & 42 & 41 & 39 & 47 & 48 & 39 & 45 & 44 & 43 & 94.2 & 92.8 & 96.7 \\
YZsel & -1 & 14 & 15 & 55 & 52 & 50 & 55 & 54 & 52 & 55 & 52 & 51 & 94.7 & 94.3 & 93.7 \\
OP + All & 13 & 22 & 9 & 34 & 33 & 33 & 36 & 40 & 34 & 34 & 33 & 33 & 93.7 & 89.7 & 93.6 \\
OP + Ysel & 16 & 22 & 6 & 34 & 33 & 33 & 38 & 40 & 33 & 35 & 33 & 33 & 93.3 & 89.8 & 94.3 \\
OP + YZsel & 16 & 21 & 5 & 34 & 33 & 33 & 38 & 39 & 33 & 35 & 33 & 33 & 92.8 & 90.5 & 94.9 \\
\multicolumn{1}{l}{\textbf{Random Forests}} &  &  &  &  &  &  &  &  &  &  &  &  &  &  &  \\
Confounder & 29 & 26 & -3 & 42 & 41 & 39 & 51 & 48 & 39 & 44 & 43 & 43 & 91.2 & 91.1 & 96.1 \\
Treatment & 51 & 38 & -13 & 39 & 38 & 37 & 64 & 54 & 39 & 42 & 41 & 41 & 78.6 & 86.2 & 96.0 \\
Outcome & 43 & 34 & -9 & 36 & 35 & 34 & 56 & 49 & 35 & 41 & 40 & 41 & 85.2 & 90.3 & 97.4 \\
All & 99 & 55 & -45 & 37 & 37 & 36 & 106 & 66 & 57 & 39 & 39 & 39 & 26.1 & 71.3 & 81.0 \\
Ysel & 54 & 38 & -16 & 36 & 35 & 34 & 65 & 52 & 38 & 40 & 40 & 40 & 75.2 & 87.3 & 96.5 \\
YZsel & 30 & 27 & -4 & 47 & 44 & 43 & 56 & 51 & 43 & 46 & 44 & 44 & 90.4 & 91.7 & 96.5 \\
OP + All & 11 & 21 & 10 & 34 & 34 & 33 & 36 & 40 & 34 & 34 & 33 & 33 & 94.2 & 89.4 & 92.7 \\
OP + Ysel & 14 & 22 & 7 & 34 & 34 & 33 & 37 & 40 & 33 & 35 & 33 & 33 & 93.3 & 90.0 & 93.7 \\
OP + YZsel & 15 & 20 & 5 & 34 & 33 & 33 & 37 & 39 & 33 & 35 & 33 & 33 & 93.2 & 90.5 & 94.6 \\ \bottomrule
\end{tabular}%
}
\caption[Simulation results for the scenario with linear main effects and linear interactions for sample size of 1000.]{Simulation results for the scenario with linear main effects and linear interactions. Each simulated dataset contained 1000 subjects. Standard errors were estimated based on 200 bootstrap replications. Results were obtained using 2000 simulated datasets.}
\label{tab:Chap3_Suppl_L-L_n1000}
\end{table}

\begin{table}[]
\resizebox{\textwidth}{!}{%
\begin{tabular}{@{}rccccccccccccccc@{}}
\toprule
\multicolumn{1}{l}{} & \multicolumn{3}{c}{Bias$\times 1000$} & \multicolumn{3}{c}{MCSD$\times 1000$} & \multicolumn{3}{c}{RMSE$\times 1000$} & \multicolumn{3}{c}{SE$\times 1000$} & \multicolumn{3}{c}{Coverage (\%; modified)} \\
\multicolumn{1}{c}{Estimators} & \multicolumn{1}{l}{1 vs 2} & 1 vs 3 & 2 vs 3 & 1 vs 2 & 1 vs 3 & 2 vs 3 & 1 vs 2 & 1 vs 3 & 2 vs 3 & 1 vs 2 & 1 vs 3 & 2 vs 3 & 1 vs 2 & 1 vs 3 & 2 vs 3 \\ \midrule
\multicolumn{1}{l}{\textbf{Naive}} & 33 & 16 & -18 & 54 & 53 & 53 & 64 & 56 & 55 & 55 & 54 & 53 & 90.5 & 94.6 & 93.8 \\
\multicolumn{1}{l}{\textbf{OAL}} & 10 & 4 & -6 & 48 & 47 & 45 & 49 & 47 & 45 & 56 & 55 & 54 & 97.6 & 98.2 & 97.8 \\
\multicolumn{1}{l}{\textbf{LOGIS}} &  &  &  &  &  &  &  &  &  &  &  &  &  &  &  \\
Confounder & 9 & 9 & -1 & 51 & 50 & 48 & 51 & 51 & 48 & 51 & 51 & 49 & 94.9 & 95.3 & 95.2 \\
Treatment & -1 & 1 & 2 & 51 & 51 & 49 & 51 & 51 & 49 & 53 & 52 & 50 & 95.0 & 95.3 & 95.1 \\
Outcome & -1 & 2 & 2 & 48 & 46 & 45 & 48 & 46 & 45 & 49 & 48 & 47 & 95.1 & 96.2 & 95.6 \\
\bottomrule
\end{tabular}%
}
\caption[Simulation results for the scenario with nonlinear main effects and linear interactions for sample size of 500.]{Simulation results for the scenario with nonlinear main effects and linear interactions. Each simulated dataset contained 500 subjects. Standard errors were estimated based on 200 bootstrap replications. Results were obtained using 2000 simulated datasets.}
\label{tab:Chap3_Suppl_NL-L_n500}
\end{table}

\begin{table}[h]
\resizebox{\textwidth}{!}{%
\begin{tabular}{@{}rccccccccccccccc@{}}
\toprule
\multicolumn{1}{l}{} & \multicolumn{3}{c}{Bias$\times 1000$} & \multicolumn{3}{c}{MCSD$\times 1000$} & \multicolumn{3}{c}{RMSE$\times 1000$} & \multicolumn{3}{c}{SE$\times 1000$} & \multicolumn{3}{c}{Coverage (\%; modified)} \\
\multicolumn{1}{c}{Estimators} & \multicolumn{1}{l}{1 vs 2} & 1 vs 3 & 2 vs 3 & 1 vs 2 & 1 vs 3 & 2 vs 3 & 1 vs 2 & 1 vs 3 & 2 vs 3 & 1 vs 2 & 1 vs 3 & 2 vs 3 & 1 vs 2 & 1 vs 3 & 2 vs 3 \\ \midrule
\multicolumn{1}{l}{\textbf{Naive}} & 36 & 17 & -19 & 39 & 39 & 38 & 53 & 42 & 42 & 39 & 38 & 38 & 83.9 & 92.8 & 92.8 \\
\multicolumn{1}{l}{\textbf{OAL}} & 6 & 5 & -1 & 34 & 34 & 33 & 34 & 34 & 33 & 39 & 39 & 38 & 97.3 & 97.5 & 97.2 \\
\multicolumn{1}{l}{\textbf{LOGIS}} &  &  &  &  &  &  &  &  &  &  &  &  &  &  &  \\
Confounder & 12 & 11 & -1 & 36 & 36 & 34 & 38 & 38 & 34 & 36 & 36 & 34 & 94.2 & 93.4 & 94.5 \\
Treatment & 2 & 4 & 2 & 37 & 36 & 34 & 37 & 36 & 34 & 36 & 36 & 34 & 95.0 & 94.6 & 94.2 \\
Outcome & 1 & 4 & 2 & 34 & 33 & 33 & 34 & 34 & 33 & 34 & 33 & 32 & 95.1 & 94.0 & 93.7 \\
\bottomrule
\end{tabular}%
}
\caption[Simulation results for the scenario with nonlinear main effects and linear interactions for sample size of 1000.]{Simulation results for the scenario with nonlinear main effects and linear interactions. Each simulated dataset contained 1000 subjects. Standard errors were estimated based on 200 bootstrap replications. Results were obtained using 2000 simulated datasets.}
\label{tab:Chap3_Suppl_NL-L_n1000}
\end{table}

\begin{table}[h]
\resizebox{\textwidth}{!}{%
\begin{tabular}{@{}rccccccccccccccc@{}}
\toprule
\multicolumn{1}{l}{} & \multicolumn{3}{c}{Bias$\times 1000$} & \multicolumn{3}{c}{MCSD$\times 1000$} & \multicolumn{3}{c}{RMSE$\times 1000$} & \multicolumn{3}{c}{SE$\times 1000$} & \multicolumn{3}{c}{Coverage (\%; modified)} \\
\multicolumn{1}{c}{Estimators} & \multicolumn{1}{l}{1 vs 2} & 1 vs 3 & 2 vs 3 & 1 vs 2 & 1 vs 3 & 2 vs 3 & 1 vs 2 & 1 vs 3 & 2 vs 3 & 1 vs 2 & 1 vs 3 & 2 vs 3 & 1 vs 2 & 1 vs 3 & 2 vs 3 \\ \midrule
\multicolumn{1}{l}{\textbf{Naive}} & 62 & 38 & -24 & 56 & 54 & 54 & 84 & 67 & 59 & 55 & 54 & 54 & 79.0 & 88.6 & 92.6 \\
\multicolumn{1}{l}{\textbf{OAL}} & 13 & 6 & -7 & 50 & 47 & 48 & 52 & 48 & 48 & 56 & 55 & 55 & 96.1 & 97.2 & 97.3 \\
\multicolumn{1}{l}{\textbf{LOGIS}} &  &  &  &  &  &  &  &  &  &  &  &  &  &  &  \\
Confounder & 9 & 8 & -1 & 52 & 51 & 50 & 53 & 51 & 50 & 52 & 50 & 50 & 94.2 & 94.0 & 94.7 \\
Treatment & 0 & 1 & 1 & 52 & 51 & 51 & 52 & 51 & 51 & 53 & 51 & 51 & 94.7 & 93.9 & 94.5 \\
Outcome & 0 & 2 & 2 & 49 & 47 & 48 & 49 & 47 & 48 & 50 & 47 & 48 & 94.0 & 94.3 & 94.6 \\
\bottomrule
\end{tabular}%
}
\caption[Simulation results for the scenario with nonlinear main effects and nonlinear interactions for sample size of 500.]{Simulation results for the scenario with nonlinear main effects and nonlinear interactions. Each simulated dataset contained 500 subjects. Standard errors were estimated based on 200 bootstrap replications. Results were obtained using 2000 simulated datasets.}
\label{tab:Chap3_Suppl_NL-NL_n500}
\end{table}

\begin{table}[h]
\resizebox{\textwidth}{!}{%
\begin{tabular}{@{}rccccccccccccccc@{}}
\toprule
\multicolumn{1}{l}{} & \multicolumn{3}{c}{Bias$\times 1000$} & \multicolumn{3}{c}{MCSD$\times 1000$} & \multicolumn{3}{c}{RMSE$\times 1000$} & \multicolumn{3}{c}{SE$\times 1000$} & \multicolumn{3}{c}{Coverage (\%; modified)} \\
\multicolumn{1}{c}{Estimators} & \multicolumn{1}{l}{1 vs 2} & 1 vs 3 & 2 vs 3 & 1 vs 2 & 1 vs 3 & 2 vs 3 & 1 vs 2 & 1 vs 3 & 2 vs 3 & 1 vs 2 & 1 vs 3 & 2 vs 3 & 1 vs 2 & 1 vs 3 & 2 vs 3 \\ \midrule
\multicolumn{1}{l}{\textbf{Naive}} & 61 & 38 & -23 & 39 & 38 & 39 & 73 & 54 & 45 & 39 & 38 & 38 & 64.6 & 82.8 & 89.3 \\
\multicolumn{1}{l}{\textbf{OAL}} & 3 & 2 & -1 & 35 & 33 & 33 & 35 & 33 & 33 & 39 & 39 & 38 & 97.2 & 97.4 & 97.4 \\
\multicolumn{1}{l}{\textbf{LOGIS}} &  &  &  &  &  &  &  &  &  &  &  &  &  &  &  \\
Confounder & 8 & 7 & -1 & 37 & 35 & 36 & 38 & 36 & 36 & 36 & 35 & 35 & 93.8 & 94.3 & 94.8 \\
Treatment & 0 & 1 & 2 & 37 & 36 & 35 & 37 & 36 & 35 & 37 & 35 & 35 & 95.0 & 94.4 & 95.1 \\
Outcome & -1 & 1 & 1 & 35 & 33 & 33 & 35 & 33 & 33 & 34 & 33 & 33 & 94.6 & 95.2 & 94.3 \\
\bottomrule
\end{tabular}%
}
\caption[Simulation results for the scenario with nonlinear main effects and nonlinear interactions for sample size of 1000.]{Simulation results for the scenario with nonlinear main effects and nonlinear interactions. Each simulated dataset contained 1000 subjects. Standard errors were estimated based on 200 bootstrap replications. Results were obtained using 2000 simulated datasets.}
\label{tab:Chap3_Suppl_NL-NL_n1000}
\end{table}

\begin{table}[h]
\resizebox{\textwidth}{!}{%
\begin{tabular}{@{}rccccccccccccccc@{}}
\toprule
\multicolumn{1}{l}{} & \multicolumn{3}{c}{Bias$\times 1000$} & \multicolumn{3}{c}{MCSD$\times 1000$} & \multicolumn{3}{c}{RMSE$\times 1000$} & \multicolumn{3}{c}{SE$\times 1000$} & \multicolumn{3}{c}{Coverage (\%; modified)} \\ \cmidrule(lr){2-4} \cmidrule(lr){5-7} \cmidrule(lr){8-10} \cmidrule(lr){11-13} \cmidrule(lr){14-16}
\multicolumn{1}{c}{Estimators} & \multicolumn{1}{l}{1 vs 2} & 1 vs 3 & 2 vs 3 & 1 vs 2 & 1 vs 3 & 2 vs 3 & 1 vs 2 & 1 vs 3 & 2 vs 3 & 1 vs 2 & 1 vs 3 & 2 vs 3 & 1 vs 2 & 1 vs 3 & 2 vs 3 \\ \midrule
\multicolumn{1}{l}{\textbf{Naive}} & -217 & -38 & 179 & 39 & 46 & 45 & 220 & 60 & 185 & 37 & 38 & 36 & 0 & 78.6 & 0.5 \\
\multicolumn{1}{l}{\textbf{OAL}} & -24 & -14 & 10 & 40 & 47 & 40 & 47 & 49 & 41 & 38 & 46 & 41 & 89 & 93.6 & 95.2 \\
\multicolumn{1}{l}{\textbf{LOGIS}} &  &  &  &  &  &  &  &  &  &  &  &  &  &  &  \\
Confounder & -3 & 3 & 6 & 36 & 46 & 42 & 36 & 46 & 42 & 36 & 46 & 42 & 94.7 & 95.1 & 94.8 \\
Treatment & -3 & 4 & 7 & 46 & 53 & 44 & 46 & 53 & 45 & 43 & 52 & 44 & 93.3 & 94.6 & 94.7 \\
Outcome & -3 & 2 & 5 & 34 & 43 & 40 & 34 & 43 & 40 & 34 & 43 & 40 & 94.5 & 95.0 & 95.2 \\
\bottomrule
\end{tabular}%
}
\caption[Simulation results for setting with censored observations.]{Simulation results for setting with censored observations. Standard errors were estimated based on 200 bootstrap replications. Results were obtained using 2000 simulated datasets.}
\label{tab:Chap3_Suppl_censor_n1000}
\end{table}

\begin{landscape}
\begin{table}[p]
\centering
{\scriptsize%
%
%
}
\caption{Number of Phecodes selected for each group of disease for 180-day risk of ER visits.}
\label{tab:Chap3_Suppl_selected_phecodes_ER_180}
\end{table}
\end{landscape}

\begin{landscape}
\begin{table}[p]
\centering
{\scriptsize%
%
%
}
\caption{Number of Phecodes selected for each group of disease for 360-day risk of ER visits.}
\label{tab:Chap3_Suppl_selected_phecodes_ER_360}
\end{table}
\end{landscape}

\begin{landscape}
\begin{table}[p]
\centering
{\scriptsize
%
%
}
\caption{Number of Phecodes selected for each group of disease for 180-day risk of hospitalization.}
\label{tab:Chap3_Suppl_selected_phecodes_hosp_180}
\end{table}
\end{landscape}

\begin{landscape}
\begin{table}[p]
\centering
{\scriptsize
%
 \\
Neurological & 46 & 5 & 2 & 0 &  \\
Respiratory & 64 & 7 & 7 & 2 & Abnormal findings examination of lungs, pneumonia \\
Sense organ & 83 & 2 & 2 & 0 &  \\
Symptoms & 36 & 6 & 6 & 2 & Malaise and fatigue, nausea and vomiting. \\ \bottomrule
\end{tabular}%
}
\caption{Number of Phecodes selected for each group of disease for 360-day risk of hospitalization.}
\label{tab:Chap3_Suppl_selected_phecodes_hosp_360}
\end{table}
\end{landscape}

\subsection*{Supplementary Figures}

\begin{figure}[h]
    \centering
    \includegraphics[width=\textwidth]{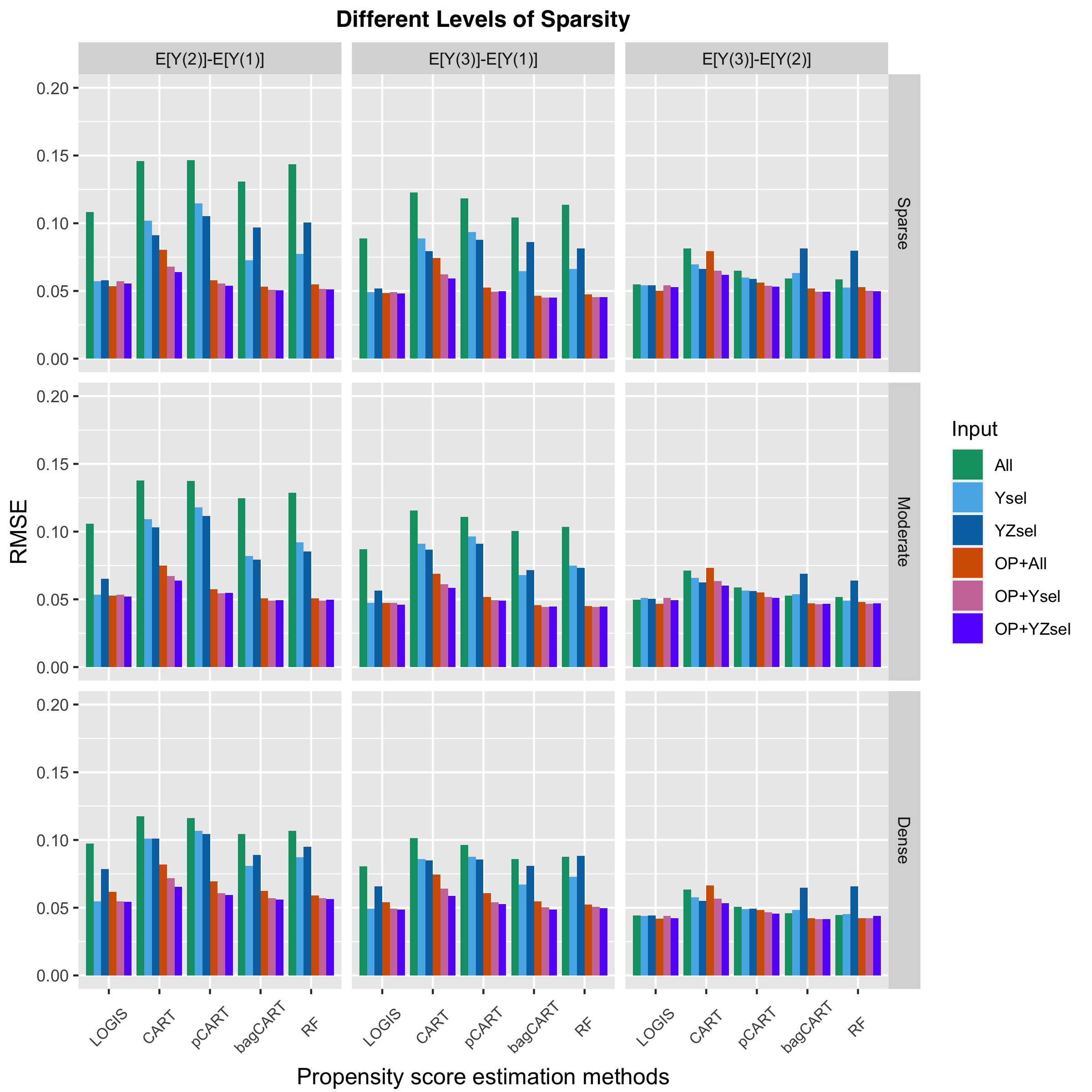}
    \caption[RMSE for 2000 inverse probability weighted estimates for the average treatment effects under scenarios with different levels of sparsity.]{RMSE for 2000 inverse probability weighted estimates for the average treatment effects under scenarios with different levels of sparsity. The rows represent scenarios and columns represent treatment pairs. Each simulated dataset contained 500 samples.}
    \label{fig:Chap3_Suppl_RMSE_sparsity}
\end{figure}

\begin{figure}[h]
    \centering
    \includegraphics[width=\textwidth]{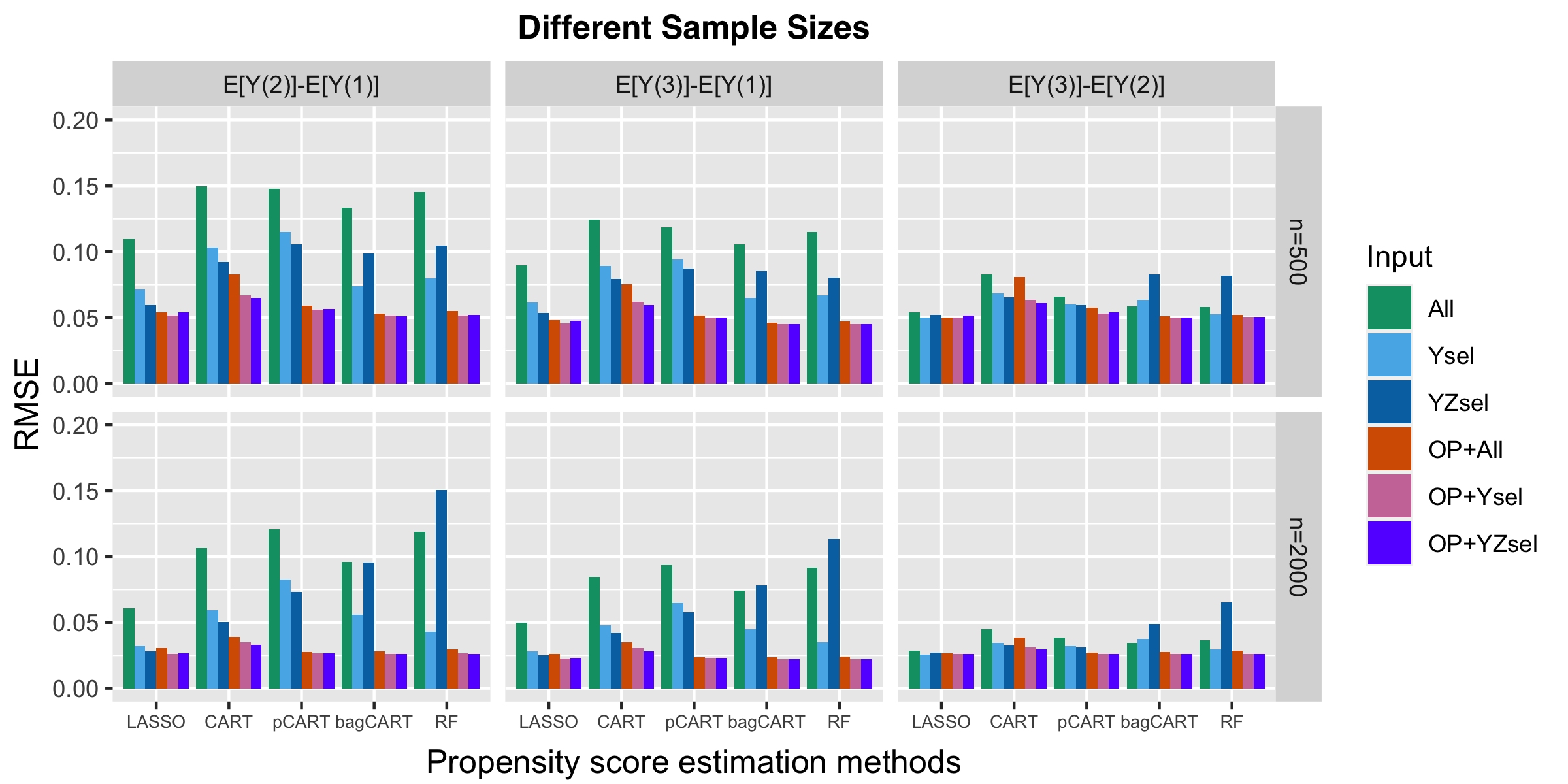}
    \caption[RMSE for 2000 inverse probability weighted estimates for the ATE for different sample sizes.]{RMSE for 2000 inverse probability weighted estimates for the ATE for different sample sizes. The rows represent scenarios and columns represent treatment pairs.}
    \label{fig:Chap3_Suppl_RMSE_sample_size}
\end{figure}

\begin{figure}[h]
    \centering
    \includegraphics[width=\textwidth]{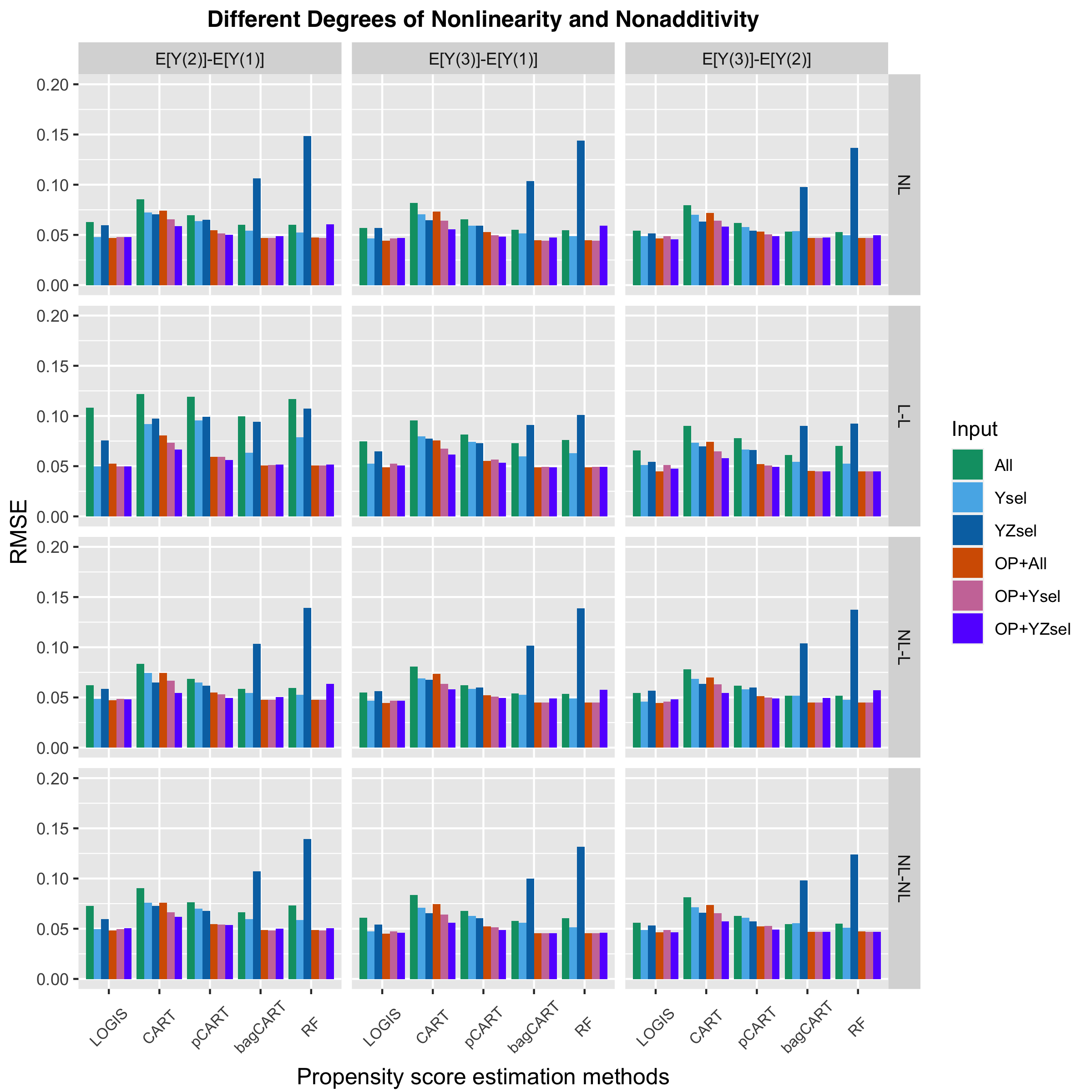}
    \caption[RMSE for 2000 inverse probability weighted estimates for the ATE under scenarios with various degrees of nonlinearity and nonadditivity in the treatment generating model.]{RMSE for 2000 inverse probability weighted estimates for the ATE under scenarios with various degrees of nonlinearity and nonadditivity in the treatment generating model. The rows represent scenarios and columns represent treatment pairs. Each simulated dataset contained 500 samples.}
    \label{fig:Chap3_Suppl_RMSE_nonlinear}
\end{figure}

\begin{figure}[h]
    \centering
    \includegraphics[width=\textwidth]{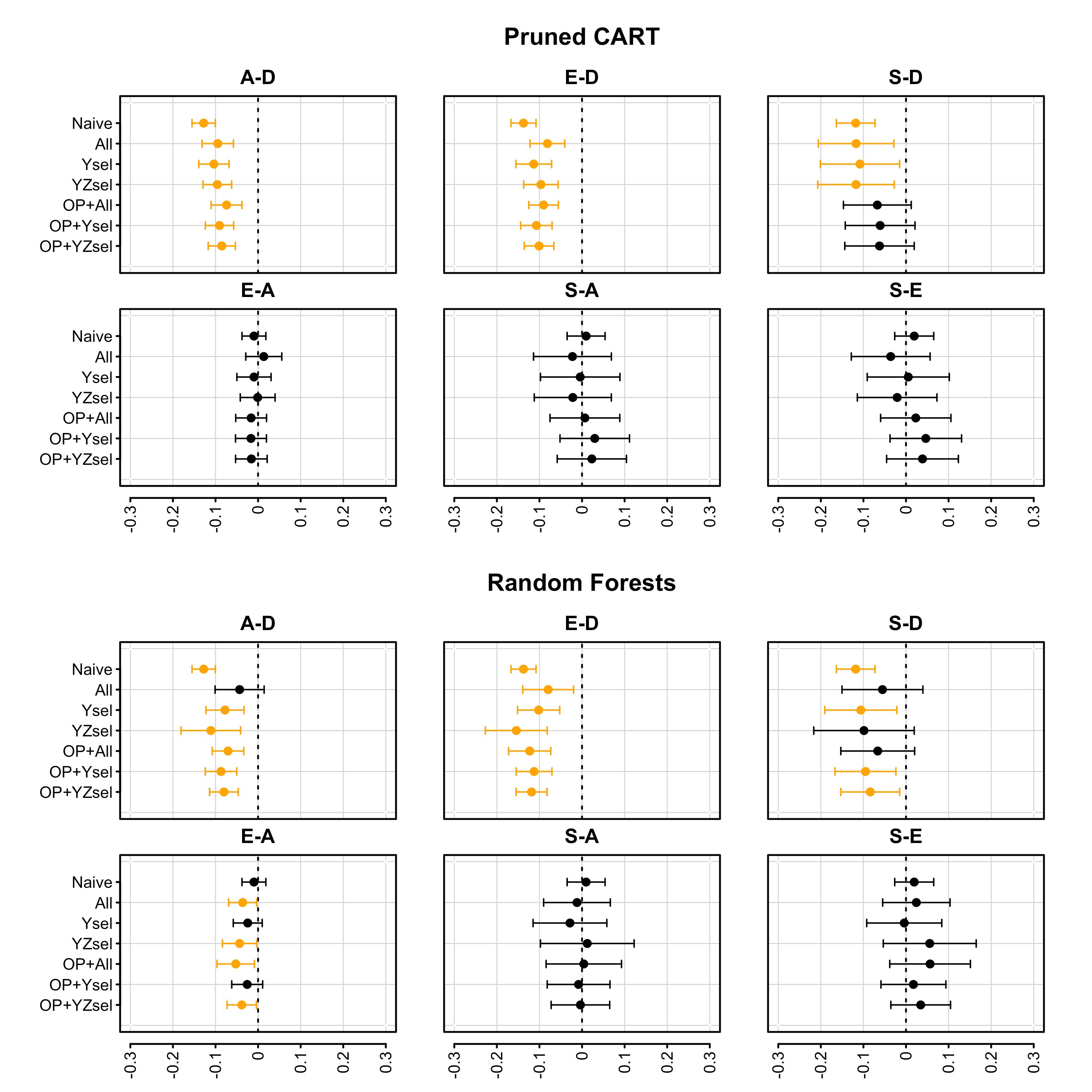}
    \caption[Average treatment effects for ER visits within 180 days of treatment initiation for pruned CART and random forests.]{Average treatment effects for ER visits within 180 days of treatment initiation for pruned CART and random forests. Data were obtained from Optum Clinformative Data Mart. Total sample size was $N=7678$ ($N_A=2757$, $N_D=2311$, $N_E=2043$, $N_S=567$). Confidence intervals that exclude zero are highlighted in orange. Abbreviations: A, abiraterone; D, docetaxel; E, enzalutamide; S, sipuleucel-T.}
    \label{fig:Chap3_ER_d180_merged2}
\end{figure}

\begin{figure}[h]
    \centering
    \includegraphics[width=0.9\textwidth]{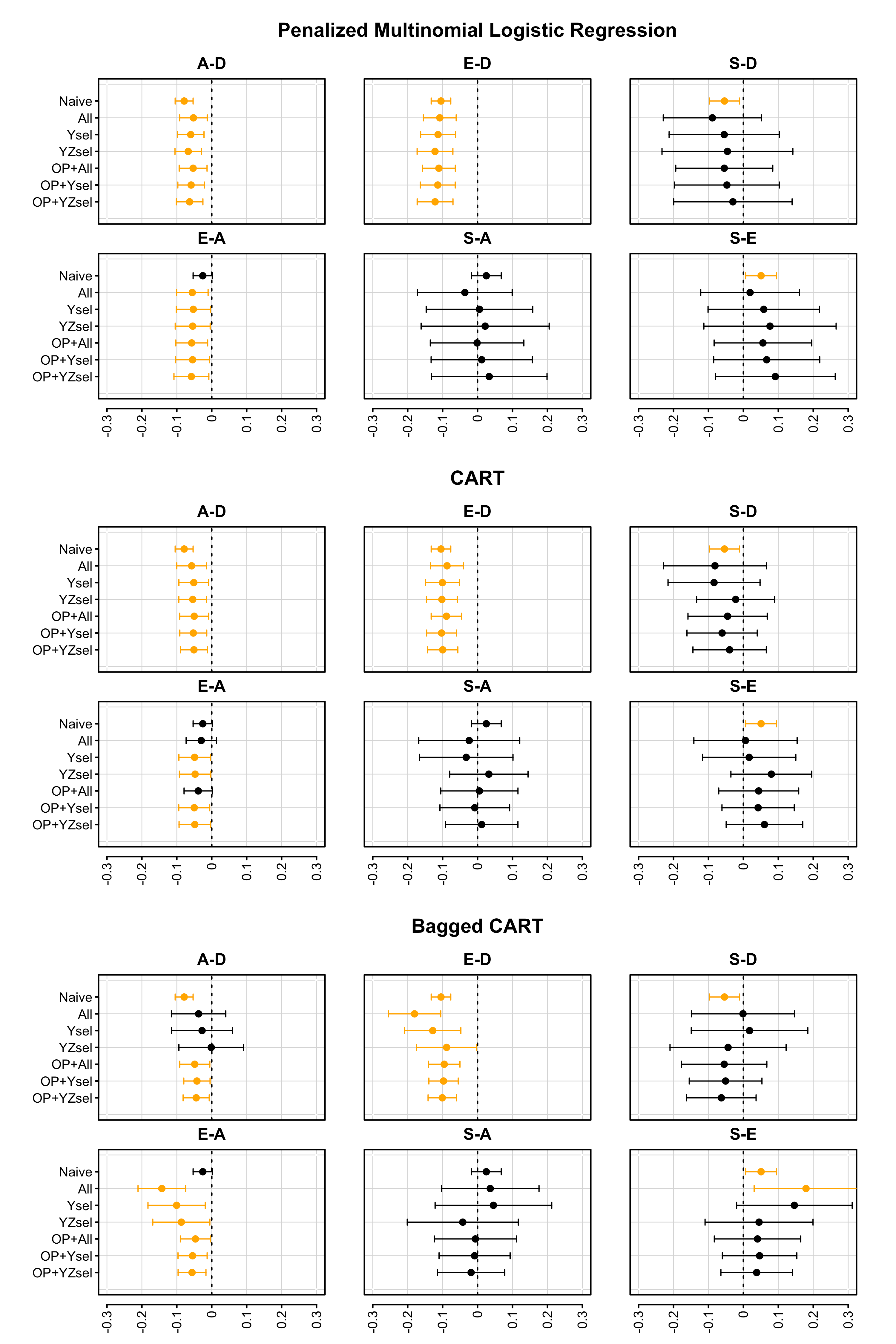}
    \caption[Average treatment effects for ER visits within 360 days of treatment initiation for LOGIS, CART and bagged CART.]{Average treatment effects for ER visits within 360 days of treatment initiation for LOGIS, CART and bagged CART. Data were obtained from Optum Clinformative Data Mart. Total sample size was $N=7678$ ($N_A=2757$, $N_D=2311$, $N_E=2043$, $N_S=567$). Confidence intervals that exclude zero are highlighted in orange. Abbreviations: A, abiraterone; D, docetaxel; E, enzalutamide; S, sipuleucel-T.}
    \label{fig:Chap3_ER_d360_merged1}
\end{figure}

\begin{figure}[h]
    \centering
    \includegraphics[width=\textwidth]{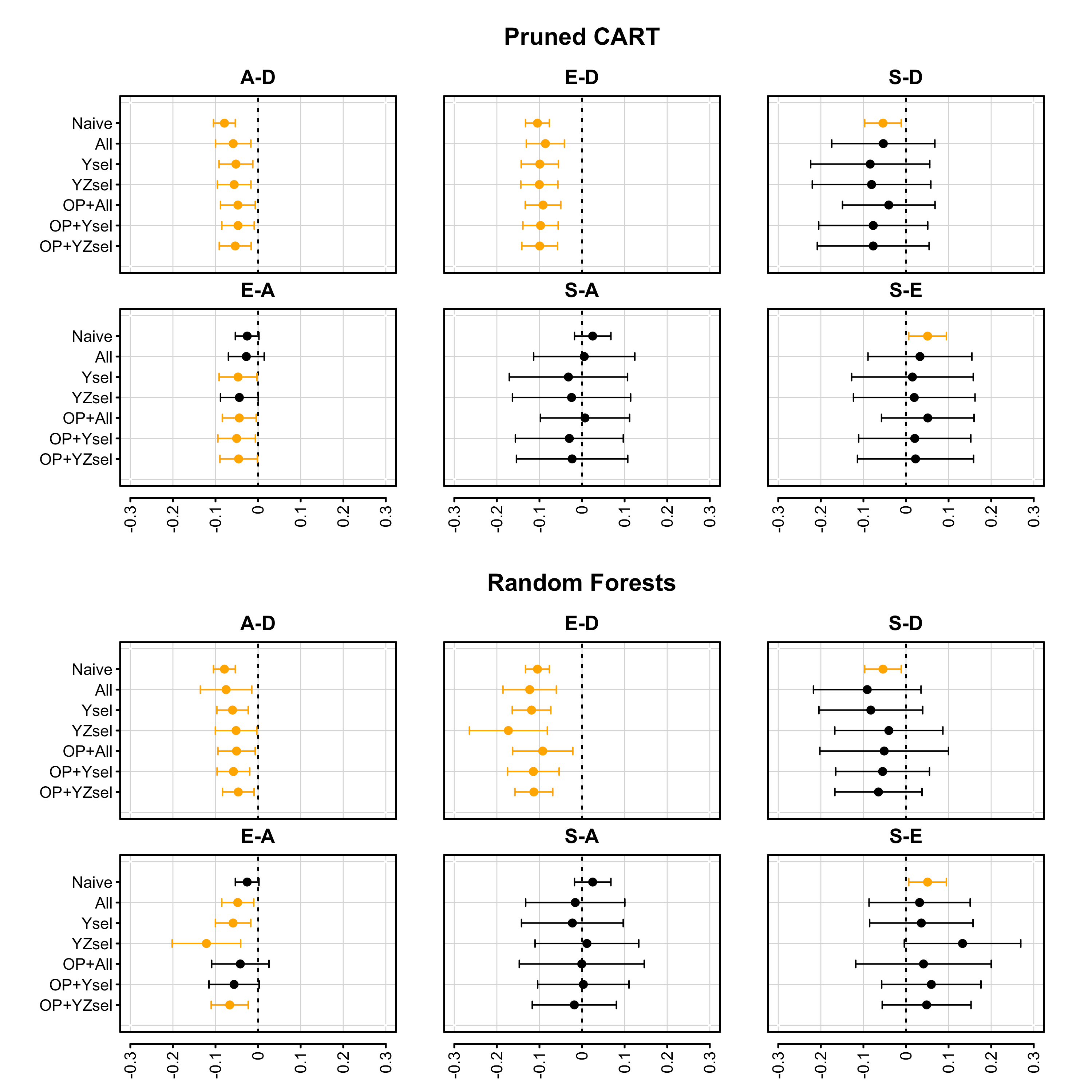}
    \caption[Average treatment effects for ER visits within 360 days of treatment initiation for pruned CART and random forests.]{Average treatment effects for ER visits within 360 days of treatment initiation for pruned CART and random forests. Data were obtained from Optum Clinformative Data Mart. Total sample size was $N=7678$ ($N_A=2757$, $N_D=2311$, $N_E=2043$, $N_S=567$). Confidence intervals that exclude zero are highlighted in orange. Abbreviations: A, abiraterone; D, docetaxel; E, enzalutamide; S, sipuleucel-T.}
    \label{fig:Chap3_ER_d360_merged2}
\end{figure}

\begin{figure}[h]
    \centering
    \includegraphics[width=\textwidth]{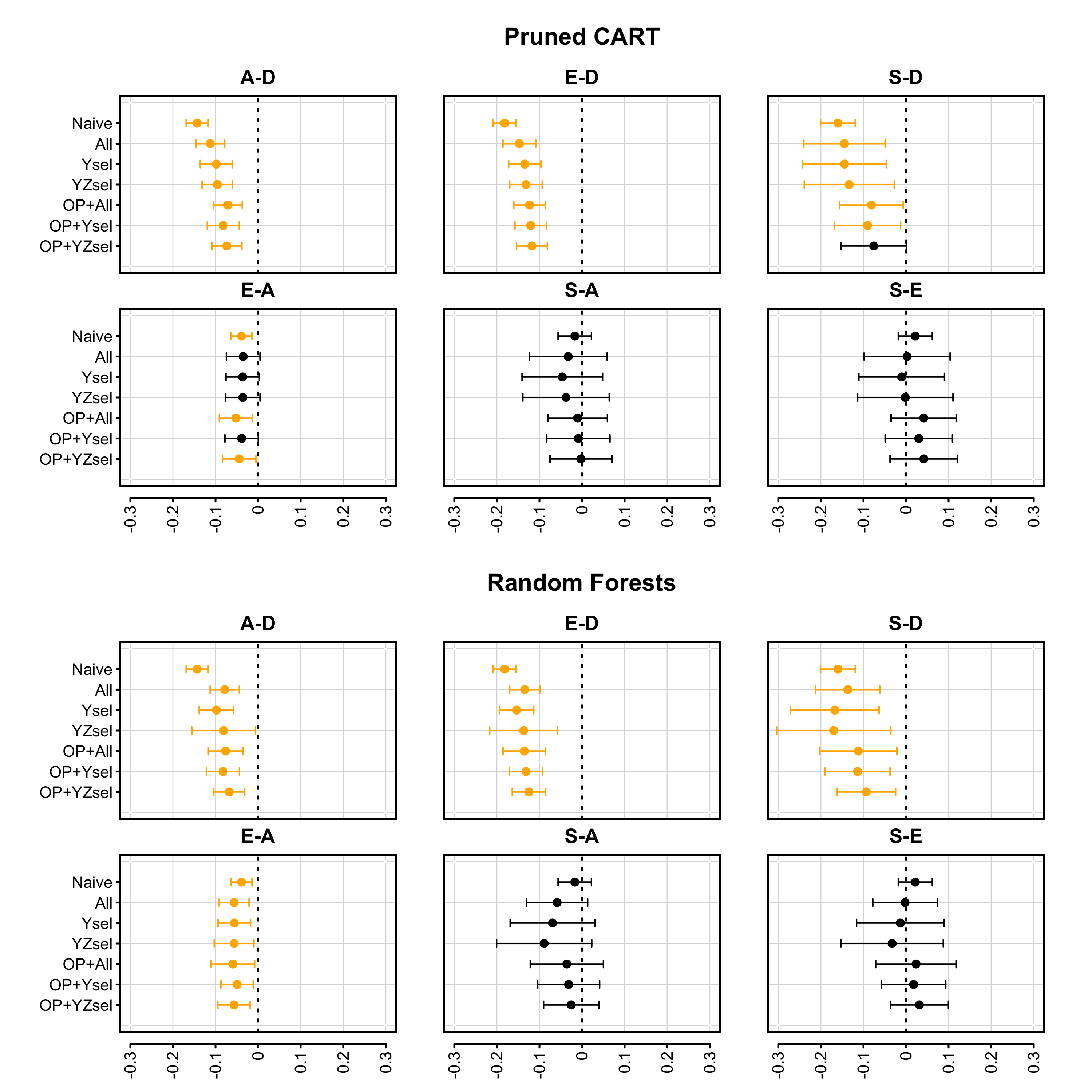}
    \caption[Average treatment effects for hospitalization within 180 days of treatment initiation for pruned CART and random forests.]{Average treatment effects for hospitalization within 180 days of treatment initiation for pruned CART and random forests. Data were obtained from Optum Clinformative Data Mart. Total sample size was $N=7709$ ($N_A=2766$, $N_D=2320$, $N_E=2051$, $N_S=572$). Confidence intervals that exclude zero are highlighted in orange. Abbreviations: A, abiraterone; D, docetaxel; E, enzalutamide; S, sipuleucel-T.}
    \label{fig:Chap3_hosp_d180_merged2}
\end{figure}

\begin{figure}[h]
    \centering
    \includegraphics[width=\textwidth]{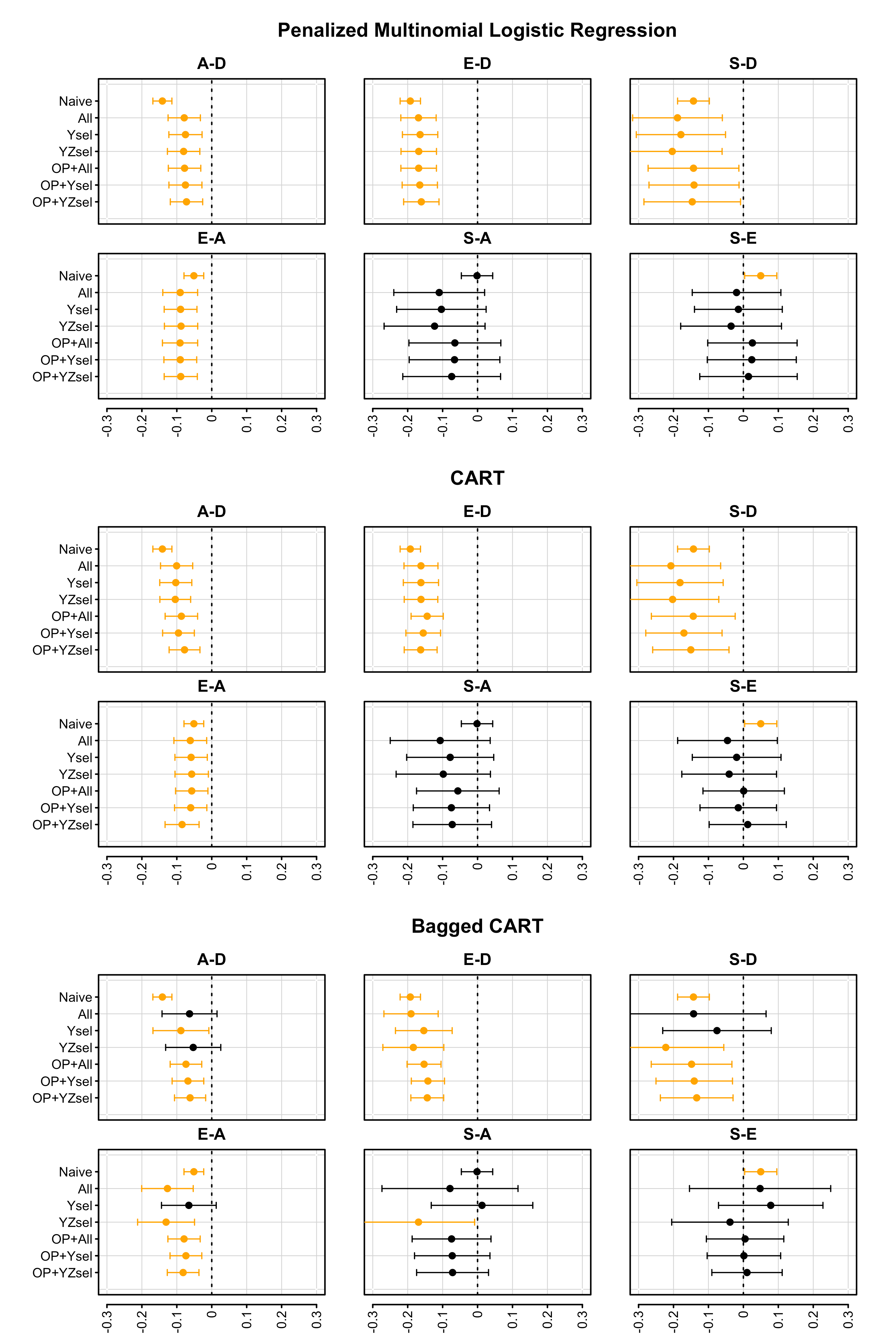}
    \caption[Average treatment effects for hospitalization within 360 days of treatment initiation for LOGIS, CART, and bagged CART.]{Average treatment effects for hospitalization within 360 days of treatment initiation for LOGIS, CART, and bagged CART. Data were obtained from Optum Clinformative Data Mart. Total sample size was $N=7709$ ($N_A=2766$, $N_D=2320$, $N_E=2051$, $N_S=572$). Confidence intervals that exclude zero are highlighted in orange. Abbreviations: A, abiraterone; D, docetaxel; E, enzalutamide; S, sipuleucel-T.}
    \label{fig:Chap3_hosp_d360_merged1}
\end{figure}

\begin{figure}[h]
    \centering
    \includegraphics[width=\textwidth]{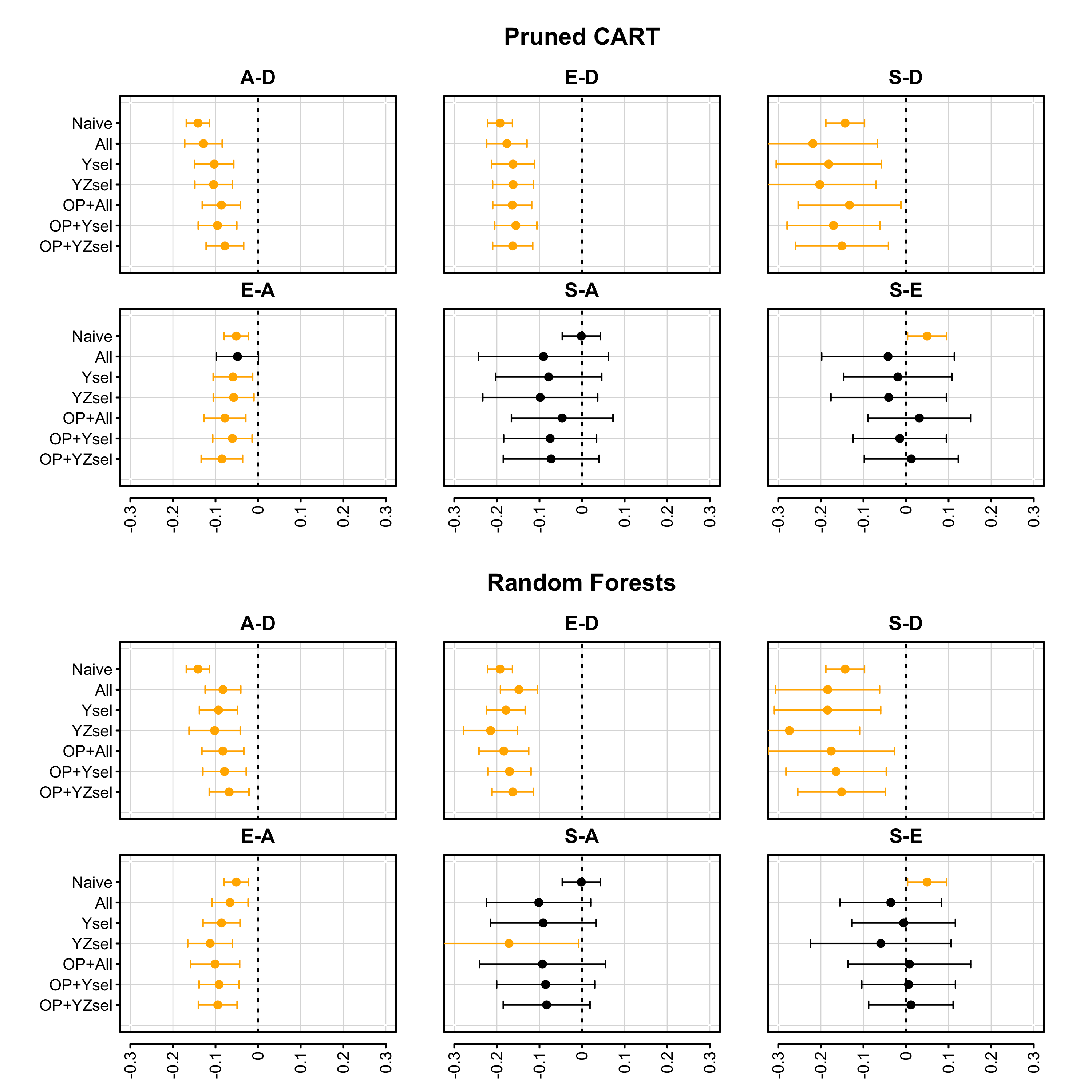}
    \caption[Average treatment effects for hospitalization within 360 days of treatment initiation for pruned CART and random forests.]{Average treatment effects for hospitalization within 360 days of treatment initiation for pruned CART and random forests. Data were obtained from Optum Clinformative Data Mart. Total sample size was $N=7709$ ($N_A=2766$, $N_D=2320$, $N_E=2051$, $N_S=572$). Confidence intervals that exclude zero are highlighted in orange. Abbreviations: A, abiraterone; D, docetaxel; E, enzalutamide; S, sipuleucel-T.}
    \label{fig:Chap3_hosp_d360_merged2}
\end{figure}

\end{document}